\documentclass[letterpaper,titlepage,11pt]{article}
\usepackage{etex}

\usepackage{latexsym}

\usepackage{amssymb}
\usepackage{amscd}
\usepackage{bbm}
\usepackage{fancybox}
\usepackage{cite}
\usepackage{amsmath,amsfonts,amsbsy}
\usepackage{pstricks,pst-node}
\usepackage{graphicx}
\usepackage{epsfig}
\usepackage{psfrag}
\usepackage{comment}
\usepackage{epstopdf}
 \usepackage{relsize}
 \usepackage{young}
\usepackage{caption}
\usepackage{hhline}

\usepackage[
      colorlinks=true,
      linkcolor=blue,
      urlcolor=black,
      filecolor=green,
      citecolor=red,
      pdfstartview=FitV,
      pdftitle={},
        pdfauthor={Hamid R. Afshar, Eric A.~Bergshoeff, Aditya Mehra, Pulastya Parekh, Blaise Rollier},
        pdfsubject={},
        pdfkeywords={},
        pdfpagemode=None,
        bookmarksopen=true
      ]{hyperref}

\usepackage[all]{hypcap}

\def\clock{{\count0=\time
           \divide\count0 60
           \ifnum\count0<10 0\fi\the\count0
           \multiply\count0 -60 \advance\count0 \time
           :\ifnum\count0<10 0\fi \the\count0
         }}
\newcommand{\timestamp}{{\small\vbox{\hbox{\tt\jobname.tex}
\hbox{\the\day/\the\month/\the\year, \clock}}}}

\psset{unit=1.3cm,linewidth=.5pt,radius=.2}  

\usepackage{multirow}                     
\usepackage{float}                          
\usepackage{lscape}                         
\usepackage{bm}

\usepackage{empheq}

\newcommand{\cD}{{\cal D}}

\newcommand{\cR}{{\cal R}}

\newcommand{\cV}{{\cal V}}

\newcommand{\beq}{\begin{equation}}
\newcommand{\eeq}{\end{equation}}
\newcommand{\bi}{\begin{itemize}}
\newcommand{\ei}{\end{itemize}}
\newcommand{\bt}{\begin{tabular}}
\newcommand{\et}{\end{tabular}}
\newcommand{\bc}{\begin{center}}
\newcommand{\ec}{\end{center}}

\def\theequation{\arabic{section}.\arabic{equation}}

\newcommand{\g}{\mathfrak{g}}

\newcommand{\be}{\begin{equation}}
\newcommand{\ee}{\end{equation}}
\newcommand{\bea}{\begin{eqnarray}}
\newcommand{\eea}{\end{eqnarray}}
\newcommand{\ba}{\begin{array}}
\newcommand{\ea}{\end{array}}

\def\bbox{{\,\lower0.9pt\vbox{\hrule \hbox{\vrule height 0.2 cm
\hskip 0.2 cm \vrule height 0.2 cm}\hrule}\,}}
\newcommand{\dsl}{\pa \kern-0.5em /}

\newcommand{\nn}{\nonumber \\}

\def\g{\gamma }

\makeatletter \@addtoreset{equation}{section} \makeatother
\renewcommand{\theequation}{\thesection.\arabic{equation}}

\def\slashchar#1{\setbox0=\hbox{$#1$}           
   \dimen0=\wd0                                 
   \setbox1=\hbox{/} \dimen1=\wd1               
   \ifdim\dimen0>\dimen1                        
      \rlap{\hbox to \dimen0{\hfil/\hfil}}      
      #1                                        
   \else                                        
      \rlap{\hbox to \dimen1{\hfil$#1$\hfil}}   
      /                                         
   \fi}

\newcommand{\pa}{\partial}

\def\mass{\footnotesize{\textrm M}}

\newcommand{\Schbox}{\square_{\textrm{\tiny Sch}}}

\newcommand*\widefbox[1]{\fbox{\hspace{0em}#1\hspace{0em}}}
\begin{document}
\begin{titlepage}
\rightline{}{UG--15--20}
\vskip 2cm
\begin{center}
 {\LARGE \bf  {\LARGE \bf A Schr\"odinger  approach to  \\[.3truecm] Newton-Cartan and Ho\v rava-Lifshitz  gravities}}
\vskip 2cm
\centerline{\bf Hamid R. Afshar$^{1}$, Eric A.~Bergshoeff$^{2}$, Aditya Mehra$^{3}$,}
\centerline{\bf Pulastya Parekh$^{3}$  and Blaise Rollier$^{2}$}
\vskip 0.5cm
\centerline{\sl $^{1}$School of Physics, Institute for Research in Fundamental Sciences (IPM)}
\centerline{\sl P.O.Box 19395-5531, Tehran, Iran}
\vskip 0.5cm
\centerline{\sl $^{2}$Van Swinderen Institute for Particle Physics and Gravity, University of Groningen}
\centerline{\sl Nijenborgh 4, 9747 AG Groningen, The Netherlands}
\vskip 0.5cm
\centerline{\sl $^{3}$Indian Institute of Science Education and Research}
\centerline{\sl Dr Homi Bhabha Road, Pashan, Pune 411008, India}
\end{center}

\vskip 0.5cm
\centerline{\small E-mail: \tt \href{mailto:afshar@ipm.ir}{afshar@ipm.ir}, \href{mailto:E.A.Bergshoeff@rug.nl}{E.A.Bergshoeff@rug.nl}, \href{mailto:aditya.mehra@students.iiserpune.ac.in}{aditya.mehra@students.iiserpune.ac.in},}
\centerline{\small \tt \href{mailto:pulastya.parekh@students.iiserpune.ac.in}{pulastya.parekh@students.iiserpune.ac.in}, \href{mailto:B.R.Rollier@rug.nl}{B.R.Rollier@rug.nl}}

\vskip 1.6cm
\centerline{\bf Abstract} \vskip 0.2cm \noindent

We define a `non-relativistic conformal method', based on a Schr\"odinger algebra with critical exponent $z=2$, as the non-relativistic version of the relativistic conformal method. An important ingredient of this method is the occurrence of a complex compensating scalar field that transforms under both scale and central charge transformations. We apply this non-relativistic method to derive the curved space Newton-Cartan gravity equations of motion with twistless torsion. Moreover, we reproduce $z=2$ Ho\v rava-Lifshitz gravity by classifying all possible Schr\"odinger invariant scalar field theories of a complex scalar up to second order in time derivatives.

\end{titlepage}

\newpage
\setcounter{page}{1} \tableofcontents

\newpage

\section{Introduction}

General Relativity as a locally Poincar\'e invariant theory of gravity has passed many experimental tests and is very successful in describing the gravitational interactions in our world. The locally supersymmetric extension of Poincar\'e gravity is called supergravity. Although supersymmetry has not yet been detected in any of the running experiments much effort has been devoted to construct supergravity invariants of all kinds of sort, with or without matter. An extremely useful technique to construct such invariants, is the (super)conformal method \cite{Kaku:1978nz,Kaku:1978ea,Ferrara:1977ij,Kaku:1977pa} where one makes use of compensating multiplets that transform under (super)conformal transformations --- for an introduction see \cite{Freedman:2012zz}.
Gauge-fixing some of the components of the compensating multiplet, one gets rid of the redundant (super)conformal symmetries and obtains the desired (super-)Poincar\'e-invariant. One of the advantages of this method is that one can use the different (super)conformal multiplets as building blocks to construct the more complicated (super-)Poincar\'e invariants. In this work we apply a similar technique to obtain non-relativistic Galilean invariants.

In recent years, gravitational theories with non-relativistic symmetries have attracted renewed attention and have been widely studied from both theoretical and phenomenological points of views. Two famous examples of such non-relativistic theories are Newton-Cartan (NC) gravity \cite{Cartan:1923zea,Cartan:1924yea} and Ho\v{r}ava-Lifshitz (HL) gravity \cite{Horava:2009uw,Horava:2008ih}. Initially, NC gravity was developed as the generally covariant arbitrary frame reformulation of Newtonian gravity. Subsequent developments showed that NC gravity correctly describes the non-relativistic limit of General Relativity, see e.g. \cite{Dautcourt}.\footnote{The underlying symmetries of NC gravity are described by the centrally extended Galilean algebra which is called the Bargmann algebra.} In a different context, in the search for a consistent theory of quantum gravity, HL gravity has been proposed as a possible UV completion of Einstein's theory. Ho\v{r}ava's theory is based on the so-called foliation preserving diffeomorphisms instead of the full diffeomorphisms of General Relativity. Following on Ho\v{r}ava's proposal, a vast literature explored the low-energy consistency and phenomenological viability of the theory, see \cite{Mukohyama:2010xz,Sotiriou:2010wn} for reviews. While HL gravity is rather unrelated to NC gravity as a gravitational theory, it has recently been shown \cite{Hartong:2015zia} that HL gravity can be reformulated in the same geometric formulation as NC gravity, namely using NC geometry --- see e.g. \cite{Trautman,Kuenzle:1972zw,Duval:1984cj} for early works on the geometric structure of NC. More precisely, the so-called non-projectable and projectable versions of HL gravity correspond to an action made of a collection of higher-derivative invariants constructed out of the variables of NC geometry respectively with and without torsion \cite{Hartong:2015zia}.

Additional independent motivations for studying non-relativistic theories of gravity come from both the potential condensed matter applications and the developments in non-relativistic holography, initially studied in \cite{Leiva:2003kd,Balasubramanian:2008dm,Son:2008ye,Herzog:2008wg,Duval:2008jg,Kachru:2008yh,Bagchi:2009my,Taylor:2008tg}. In particular, HL gravity is interesting from the holographic point of view since it provides an alternative way of realizing a non-relativistic holographic model where the bulk and the boundary have the same non-relativistic geometric structure\cite{Janiszewski:2012nb,Griffin:2012qx}. 
This has recently been argued to be a very natural, and promising, approach in the  context of certain non-relativistic field theories called Warped CFT's \cite{Hofman:2014loa,Castro:2015csg}. In fact, many condensed matter systems are described by non-relativistic field theories and coupling these systems to non-relativistic backgrounds provides appropriate external sources conjugated to their conserved currents. In the context of NC geometry, the appropriate way of performing this coupling has been studied from the theoretical point of view \cite{Jensen:2014aia,Jensen:2014ama,Hartong:2014pma,Hartong:2015wxa,Geracie:2015dea,Geracie:2015xfa} and typical condensed matter applications include, e.g., the description of the quantum Hall effect \cite{Son:2013rqa,Gromov:2014vla,Geracie:2014nka,Moroz:2014ska}.

A crucial aspect of Newton-Cartan geometry, particularly relevant in the developments mentioned just above, is that the geometry can involve a non-vanishing torsion tensor. The Torsional Newton-Cartan (TNC) geometry was first observed in \cite{Christensen:2013rfa,Christensen:2013lma} in the context of Lifshitz holography\footnote{
See \cite{Taylor:2015glc} for a recent review on Lifshitz holography.} where it arises as the boundary geometry of a specific model supporting $z=2$ Lifshitz solutions. Following \cite{Andringa:2010it} on the gauging of the Bargmann algebra, it was later shown that the TNC geometry can be obtained by gauging the Schr\"odinger algebra \cite{Bergshoeff:2014uea}. In this work, we will exclusively consider a specific type of torsion referred to as `twistless torsion'.


The purpose of this paper is to develop a non-relativistic version of the conformal method mentioned at the beginning of this introduction and to illustrate it in two different contexts. We apply this method at the level of the equations of motion to find the curved space covariant NC gravity field equations with torsion and we apply it to re-derive $z=2$ HL gravity by reproducing the Galilean invariants constructed in \cite{Hartong:2015zia}. We expect that this procedure will be very efficient if one considers more complicated models such as supersymmetric HL or NC gravity theories. In our formalism we use Schr\"odinger gravity, i.e.~the gauge theory of the Schr\"odinger algebra \cite{Hagen:1972pd,Niederer:1972zz}, as the analogue  of conformal gravity in the relativistic case. We remind that the Schr\"odinger transformations are the maximal symmetries that leave invariant the action of a massive non-relativistic particle whereas the action of a massless non-relativistic particle is invariant under the symmetries of the so-called Galilean conformal algebra \cite{Barut:1973,Havas:1978}. We will therefore reserve the name `non-relativistic conformal gravity' for the gauge theory of the Galilean conformal algebra.\,\footnote{We thank J.~Lukierski for a discussion on this point.} For our present purposes, it is sufficient to make use of Schr\"odinger gravity.

The Schr\"ödinger symmetries in $d$ spatial dimensions with critical exponent $z$, which we denote by $Sch_z(d)$, contain in addition to the centrally extended Galilean symmetries, a dilatation  generator $D$ which acts on space and time coordinates differently. To be precise, for general exponent $z$ the  time and space coordinates transform under a dilatation with parameter $\lambda$ as follows:
\bea
t\rightarrow \lambda^z t\qquad\text{and}\qquad \text{\bf x}\rightarrow \lambda \text{\bf x}\,.
\eea
At $z=2$ an extra generator $K$, denoted as the generator of special conformal transformation,  emerges. These two extra symmetries $D$ and $K$, together with the translations (with generators  $P_a$), spatial rotations (with generators  $J_{ab}$) and  Galilean boosts (with
generators $G_a$) are symmetries
of the Schr\"odinger equation;
\bea
\left(i\partial_0-\frac{1}{2\mass}\partial_a^2\right)\Psi(t,\text{\bf x})=0\,,
\eea
where $\mass$ appears as the central term in the commutator of Galilean boosts and translations:
\bea
[P_a, G_b]=\delta_{ab} \mass\,.
\eea
The corresponding central charge transformation acts as a phase transformation  on the complex scalar $\Psi$. In the context of the Schr\"odinger equation this symmetry corresponds to particle number conservation. The field theories invariant under Schr\"odinger symmetries have been studied in  \cite{Jackiw:1990mb,Henkel:1993sg,Nishida:2007pj}.

The conformal method is based on a St\"uckelberg mechanism for a compensating scalar field involved in the conformal multiplet.
It turns out that in the non-relativistic case we  need to use a compensating complex scalar that transforms under dilatations  and central charge transformations.
This is different from the relativistic case where a real scalar is sufficient and the central charge transformations are absent. There is one more difference
with the relativistic case: while in the relativistic case the special conformal transformations are described by a vector generator $K_\mu$, in the
non-relativistic case we only have a scalar generator $K$. As a consequence of this, we cannot  gauge away the space components $b_a$  of the dilatation gauge field in the non-relativistic case. These remnants of the dilatation gauge field are precisely the origin of torsion in NC gravity \cite{Bergshoeff:2014uea}.

It is natural to apply the non-relativistic conformal method we develop in this work to NC gravity itself, in particular to obtain the extension with torsion. As far as we know,
NC gravity cannot be derived from an action principle, at least not with the usual field content. We therefore apply the non-relativistic method  at the level of the equations of motion and only consider equations of motion for the compensating scalar. It turns out that the non-relativistic conformal method in this case provides a very efficient way of constructing the equations of motion of NC gravity with torsion, a result that, as far as we know, has not occurred before  in the literature.

We also apply the non-relativistic conformal method to Schr\"odinger invariant scalar field theories. By reproducing the Galilean invariants of \cite{Hartong:2015zia}, we show that Schr\"odinger field theories (SFT's) are naturally mapped to HL gravity. Since HL gravity contains higher derivatives we need to consider higher-derivative SFT's for the compensating complex scalar. For our purposes it is sufficient to classify all SFT's up to two derivatives in time and four derivatives in space. Following the same procedure as in the relativistic case, we will couple these SFT's to Schr\"odinger gravity and next gauge-fix the compensating complex scalar after which we obtain higher-derivative Galilean invariants.

This paper is organized as follows. In section \ref{relcons} we review the relativistic conformal construction of Poincar\'e gravity.  In section \ref{nonrelcons}
we develop the non-relativistic conformal method that will be used in the remainder of the paper. We then use the non-relativistic conformal method to derive the Newton-Cartan equations of motion with and without torsion in
section \ref{NC}. In section \ref{HLgravity} we couple the complex  compensating scalar to Schr\"odinger gravity, gauge fix the scalar and thereby construct all Galilean invariants with at most two time and four spatial derivatives that can be related to a SFT. This construction leads to a systematic derivation of Ho\v rava-Lifshitz gravity at $z=2$.
Finally, we present our conclusions in section \ref{conclusion}. There are two appendices. In appendix \ref{App A} we give several details of the Schr\"odinger gravity theory, while in appendix  \ref{Sch SFT},
we classify all  scalar field theories that are invariant under rigid $z=2$ Schr\"odinger symmetries up to 2 derivatives in time and 4 derivatives in space.

Before starting we mention some notation and conventions. We work in $D = d + 1$ spacetime dimensions where $d$ refers to the number of spatial dimensions.
The small Latin alphabet letters $(a, b, c, \ldots)$ refer to the spatial local Galilean frame while
the capitals $(\text{\small A}, \text{\small B}, \text{\small C}, \ldots)$ cover the local Poincar\'e frame.
The Greek indices $(\mu, \nu, \rho, \ldots)$ refer to the coordinate frame and labels all spacetime coordinates $(x\equiv t,\text{\bf x})$.

\section{The relativistic conformal method}\label{relcons}

Before discussing the non-relativistic case, it is instructive to first review the (bosonic) relativistic conformal construction. In the relativistic conformal construction the aim is to construct general Poincar\'e invariants by using the larger conformal symmetry algebra. The idea is that conformal field theories (CFT's) of a real scalar field correspond to a class of Poincar\'e invariants. The converse is not necessarily true, see below.
\subsection{Einstein-Hilbert invariant}\label{EHinv}
We explain first the relation between the Einstein-Hilbert invariant in $D>2$ dimensions and the CFT of a free real scalar with a standard 2-derivative
kinetic term. To be explicit, we consider the Einstein-Hilbert Lagrangian in $D>2$ dimensions
\begin{align}\label{PI}
\mathrm{P}_1: \;\; e^{-1}{\cal L} = \frac{1}{\kappa^2}\,\mathcal{R}\,, \qquad
\end{align}
where  $e$ is the determinant of the Poincar\'e vielbein $e_\mu{}^{\text{\tiny A}}$ with $\text{\small A}$ and $\mu$ referring to the local Poincar\'e   and the coordinate frames respectively. The gravitational coupling constant  $\kappa^2=16\pi G$ has the length-dimension $D-2$.
Clearly, the action corresponding to this Lagrangian is dimensionless and invariant under general coordinate transformations with parameter $\xi^\mu$ and under local Lorentz transformations with parameter $\Lambda^{\text{\tiny AB}}=-\Lambda^{\text{\tiny BA}}$:
\begin{align}
\delta e_\mu{}^{\text{\tiny A}} = \xi^\lambda\partial_\lambda\, e_\mu{}^{\text{\tiny A}} + (\partial_\mu\,\xi^\lambda)\, e_\lambda{}^{\text{\tiny A}} +
\Lambda^{\text{\tiny AB}}\, e_\mu{}_{\text{\tiny B}}\,.
\end{align}

To relate the Poincar\'e invariant \eqref{PI} to a real scalar field theory we first observe that the Einstein-Hilbert action is not invariant under local dilatations.
To make it invariant under local dilatations we replace the Poincar\'e vielbein, which from now on we denote with  $(e_\mu{}^{\text{\tiny A}})^{\text{\tiny P}}$, by a compensating real scalar $\varphi$ times the conformal vielbein $(e_\mu{}^{\text{\tiny A}})^{\text{\tiny C}}$:
\begin{align}\label{rescaling}
(e_\mu{}^{\text{\tiny A}})^{\text{\tiny P}} = \kappa^{\frac{2}{D-2}}\,\varphi\,(e_\mu{}^{\text{\tiny A}})^{\text{\tiny C}}\,,
\end{align}
where $\varphi$ has a lenght-dimension -1. We inserted a factor with $\kappa$ in \eqref{rescaling} so that the dilatation invariant action will be free of any dimensionful parameter under the field redefinition \eqref{rescaling}. The compensating scalar $\varphi$ and the conformal vielbein $(e_\mu{}^{\text{\tiny A}})^{\text{\tiny C}}$ transform under dilatations, with a local parameter $\Lambda_{\text{D}}(x)$,  such that the Poincar\'e vielbein $(e_\mu{}^{\text{\tiny A}})^{\text{\tiny P}}$ is invariant:
 \begin{align}
 \delta\varphi = -\Lambda_{\text{D}}\varphi\,,\hskip 2truecm  \delta (e_\mu{}^{\text{\tiny A}})^{\text{\tiny C}} = \Lambda_{\text{D}} (e_\mu{}^{\text{\tiny A}})^{\text{\tiny C}}\,.
 \end{align}
 We next substitute the decomposition \eqref{rescaling}  into the Poincar\'e invariant \eqref{PI} and in this way end up with a real scalar coupled to conformal gravity. The corresponding action is invariant under general coordinate transformations, local Lorentz rotations and local dilatations.
 To obtain the  conformal real scalar field theory, before coupling to conformal gravity,  we impose
the gauge-fixing condition\,\footnote{Note that after gauge-fixing
 we do not distinguish between curved and flat indices anymore.}
\begin{align}\label{gfc}
(e_\mu{}^{\text{\tiny A}})^{\text{\tiny C}} = \delta_\mu{}^{\text{\tiny A}}\,.
\end{align}
This gauge-fixing condition restricts the local conformal transformations to the rigid ones via the constraint equation
\begin{align}
\partial_\mu\xi^\nu + \Lambda^{\nu}{}_{\mu} + \Lambda_{\text{D}}\delta_\mu{}^\nu = 0\,.
\end{align}
This differential equation has the following solution
\begin{subequations}
	\label{rigidconf}
\begin{align}
\xi^\mu(x) &= a^\mu - \lambda^{\mu\nu}x_\nu - \lambda_{\text{D}} x^\mu + \lambda_{\text{K}}^\nu\big(x^\mu x_\nu - \tfrac{1}{2}\delta^\mu_\nu\,x^2\big)\,,\\[.1truecm]
\Lambda^{\mu\nu}(x) &= \lambda^{\mu\nu} + 2\lambda_{\text{K}}^{[\mu}\,x^{\nu]}\,,\\[.1truecm]
\Lambda_{\text{D}}(x) &= \lambda_{\text{D}} - \lambda_{\text{K}}^\mu\, x_\mu\,,
\end{align}
\end{subequations}
where $a^\mu\,,\lambda^{\mu\nu}\,,\lambda_{\text{D}}$ and $\lambda_{\text{K}}^\mu$ are the (constant) parameters of translations, Lorentz transformations, dilatations and special conformal transformations, respectively.
The gauge-fixing condition \eqref{gfc} has the consequence that, when substituting expression \eqref{rescaling} back into \eqref{PI} one can ignore any derivative acting on the conformal vielbein. One thus ends up with a Lagrangian with  the derivatives acting on the compensating scalar $\varphi$.

Finally, we make the redefinition (assuming $D>2$)
\begin{align}
\varphi=\phi^{\frac{2}{D-2}}\label{eq: field redef}
\end{align}
such that the Lagrangian \eqref{PI} reduces to the following canonical form:\footnote{As a general feature a positive kinetic term for the compensating scalar corresponds to a negative kinetic term for gravity and vice-versa \cite{Freedman:2012zz}.}
\begin{align}\label{scalarcft}
\mathrm{CFT}_1: \;\; {\cal L} = 4\,\tfrac{D-1}{D-2}\, \phi \square \phi\,, \qquad
\end{align}
where $\square=\eta^{\mu\nu}\partial_\mu\partial_\nu$ and $\eta^{\mu\nu}$ is the inverse flat Minkowski metric. The action corresponding to this Lagrangian in $D$ dimensions is explicitly invariant under the rigid conformal transformations
\begin{align}\label{below}
\delta\phi = \xi^\mu\partial_\mu\,\phi + w\Lambda_{\text{D}}\phi\,,
\end{align}
with $\xi^\mu$ and $\Lambda_{\text{D}}$ given in \eqref{rigidconf} and, due to the redefinition \eqref{eq: field redef}, with weight $w$ given by
\begin{align}
w=-\tfrac{1}{2}(D-2)\label{weight}\,.
\end{align}

 We thus have shown how the Poincar\'e invariant P$_1$ given in \eqref{PI} can be related to the
conformal field theory CFT$_1$ of a free real scalar defined in \eqref{scalarcft}.

The relation also works the other way around. Starting from the  CFT$_1$ given in  \eqref{scalarcft} one can derive the Einstein-Hilbert Lagrangian \eqref{PI} as follows.
The first step is to make the CFT$_1$ of  \eqref{scalarcft} invariant under local conformal transformations, i.e.~couple it to conformal gravity.
In order to do this, it is convenient to  first introduce all the gauge fields of conformal gravity, not only the conformal vielbein.
By applying a standard gauging procedure to the relativistic conformal algebra, see e.g. \cite{Freedman:2012zz}
and references therein,\footnote{Note that our sign conventions are different from \cite{Freedman:2012zz}.} one arrives at the following gauge fields and transformation rules ---  we omit the superscript {\small C} from now on;
\begin{subequations}
	\label{conftransfbmu}
\begin{align}
\delta e_\mu{}^{\text{\tiny A}} &=  \Lambda^{\text{\tiny AB}}e_\mu{}^{\text{\tiny B}} + \Lambda_{\text{D}}e_\mu{}^{\text{\tiny A}}\,, \\[.1truecm]
\delta  \omega_\mu{}^{\text{\tiny AB}} &= D_\mu\Lambda^{\text{\tiny AB}} +4 \Lambda_{\text{K}}^{[\text{\tiny A}} e_\mu{}^{\text{\tiny B}]}\,, \\[.1truecm]
\delta b_\mu &= \partial_\mu \Lambda_{\text{D}} - 2\Lambda_{\text{K}}^{\text{\tiny A}} e_\mu{}_{\text{\tiny A}}\,,\label{b-transfmnt}\\[.1truecm]
\delta f_\mu{}^{\text{\tiny A}} &= D_\mu \Lambda_{\text{K}}^{\text{\tiny A}} +\Lambda^{\text{\tiny AB}}f_\mu{}_{\text{\tiny B}} +  b_\mu\Lambda_{\text{K}}^{\text{\tiny A}} -\Lambda_{\text{D}}f_\mu{}^{\text{\tiny A}}\,,
\end{align}
\end{subequations}
where $D_\mu$ denotes the Lorentz-covariant derivative and $\Lambda_{\rm K}^{\text{\tiny A}}$ is the parameter of a special conformal transformation. All gauge fields transform as covariant vectors under general coordinate transformations.

The special thing about the three new gauge fields that we have introduced is that two of them, $\omega_\mu{}^{\text{\tiny AB}}$ and $f_\mu{}^{\text{\tiny A}}$, are dependent, i.e.~$\omega_\mu{}^{\text{\tiny AB}} = \omega_\mu{}^{\text{\tiny AB}}(e,b)$ and $f_\mu{}^{\text{\tiny A}} = f_\mu{}^{\text{\tiny A}}(e,b)$,
whereas the third gauge field $b_\mu$ transform as a shift under special conformal transformations, see eq.~\eqref{b-transfmnt}. The
fact that the gauge field $b_\mu$ is the only field that transforms with a shift under the special conformal transformations\,\footnote{The gauge fields $\omega_\mu{}^{\text{\tiny AB}}(e,b)$ and $f_\mu{}^{\text{\tiny A}}(e,b)$ transform only under special
conformal transformations via their dependence on $b_\mu$.}
means that, writing out all covariant derivatives, one finds that the dilatation gauge field always drops out in any conformal invariant action in $D>2$.

Despite the fact that $b_\mu$ does not occur in the final answer, it is useful to keep this gauge field  at a first stage to couple the real scalar conformal field theory \eqref{scalarcft} to conformal gravity and construct
covariant derivatives in a systematic way. This goes as follows. Since the scalar field $\phi$ only transforms under general coordinate transformations and dilatations it's conformal covariant derivative is defined as follows:
\begin{align}\label{expression}
D_{\text{\tiny A}}\phi = e_{\text{\tiny A}}{}^\mu\big(\partial_\mu - w b_\mu\big)\phi\,.
\end{align}
Due to the presence of the dilatation gauge field $b_\mu$ this
conformal covariant derivative transforms under special conformal transformations and, therefore, if we take
the conformal covariant derivative of the expression \eqref{expression} it will involve the gauge field of special conformal transformations. The exact expression for the conformal d'Alembertian reads:
\begin{eqnarray}
\square^{\text{\tiny C}}\phi &\equiv&  \eta^{\text{\tiny AB}}D_{\text{\tiny A}}D_{\text{\tiny B}}\phi \nonumber\\[.1truecm]
&=& e^{\text{\tiny A}}{}^{\mu}\bigg(
\partial _\mu D_{\text{\tiny A}}\phi - (w-1)b_\mu D_{\text{\tiny A}}\phi -\omega_{\mu \text{\tiny A}}{}^{\text{\tiny B}}(e,b)D_{\text{\tiny B}}\phi - 2w f_{\mu \text{\tiny A}}(e,b)\phi
\bigg)\,.\label{dalemb}
\end{eqnarray}
The Lagrangian describing the coupling of $\phi$ to conformal gravity is then given by
\begin{align}\label{mattercoupled}
 e^{-1}{\cal L} = 4\,\tfrac{D-1}{D-2}\,\phi \square^{\text{\tiny C}}\phi\,.
\end{align}
The theory described by this Lagrangian is invariant under the local conformal transformations \eqref{below}, where $\xi^\mu$ and $\Lambda_{\rm D}$ are now taken to be arbitrary functions of the spacetime coordinates,  and \eqref{conftransfbmu} provided that the scalar weight $w$ is given by \eqref{weight} --- see \cite{Freedman:2012zz} for more details.

To obtain the Einstein-Hilbert action we now fix the dilatations by imposing the gauge-fixing condition
\begin{align}\label{gfc2 rel}
\phi =\frac{1}{\kappa}\,.
\end{align}
Note that this gauge-fixing condition does not require any compensating transformations. As we already mentioned, the $b_\mu$ gauge field will automatically drop out from \eqref{mattercoupled} as a consequence of the special conformal invariance. Therefore, from the expression \eqref{dalemb} it is clear that the $b_\mu$ independent part of $e_{\text{\tiny A}}{}^\mu f_\mu{}^{\text{\tiny A}}$ is the only relevant term that will ultimately survive the gauge-fixing condition \eqref{gfc2 rel}.
Using the explicit expressions for $\omega_\mu{}^{\text{\tiny AB}}$ and $f_\mu{}^{\text{\tiny A}}$,
\begin{align}
\omega_\mu{}^{\text{\tiny AB}}(e,b) &= \Omega_\mu{}^{\text{\tiny AB}}(e) + 2e_\mu{}^{[\text{\tiny A}}e^{\text{\tiny B}]\nu}b_\nu\,,\label{eq: sol rel omega}\\
f_\mu{}^{\text{\tiny A}}(e,b) &= \frac{1}{2(D-2)}\bigg(R_{\mu}{}^{\text{\tiny A}}(\omega) -\frac{1}{2(D-1)}e_\mu{}^{\text{\tiny A}}R(\omega)\bigg)\,,\label{eq: sol rel fmu}
\end{align}
with
\begin{align}
\Omega_\mu{}^{\text{\tiny AB}}(e) &= -2e^\nu{}^{[\text{\tiny A}} \partial_{[\mu} e_{\nu]}{}^{\text{\tiny B}]}+e_\mu{}_{\text{\tiny C}}e^\nu{}^{\text{\tiny A}}e^\rho{}^{\text{\tiny B}} \partial_{[\nu} e_{\rho]}{}^{\text{\tiny C}} \,,\\
R_{\mu}{}^{\text{\tiny A}}(\omega) &= 2e_{\text{\tiny B}}{}^\nu\left(\partial_{[\mu}\omega_{\nu]}{}^{\text{\tiny AB}} - \eta_{\text{\tiny CD}}\omega_{[\mu}{}^{\text{\tiny AC}}\omega_{\nu]}{}^{\text{\tiny DB}}\right)\,,
\end{align}
and $R(\omega)=e_{\text{\tiny A}}{}^\mu R_{\mu}{}^{\text{\tiny A}}(\omega)$, one finds that after gauge-fixing the matter coupled conformal gravity Lagrangian \eqref{mattercoupled} precisely reduces to the Einstein-Hilbert Lagrangian \eqref{PI} where the Poinca\'re Ricci scalar is
\begin{equation}
\mathcal{R}=2e^\mu_{\text{\tiny A}}e^\nu_{\text{\tiny B}}\left(\partial_{[\mu}\Omega_{\nu]}{}^{\text{\tiny AB}} - \eta_{\text{\tiny CD}}\Omega_{[\mu}{}^{\text{\tiny AC}}\Omega_{\nu]}{}^{\text{\tiny DB}}\right)\,.
\end{equation}

This concludes our discussion of how the Einstein-Hilbert invariant is related to the CFT of a free real scalar with a 2-derivative kinetic term.
A few remarks are in order.
First of all, we note that in the relativistic case  the  number of derivatives in the Poincar\'e-invariant
 is the same as the number of derivatives in the corresponding CFT. As we will see later on, this will no longer be true in the non-relativistic case.
Secondly, the way we couple the scalar to conformal gravity is basically by replacing derivatives by conformal-covariant derivatives.
This only works nicely if we perform the covariantization on the Lagrangian \eqref{scalarcft}. For instance, had we used as the CFT Lagrangian ${\cal L}^\prime \sim \partial_\mu\phi\partial^\mu\phi$ instead of ${\cal L} \sim \phi\square\phi$ we would not have succeeded to couple ${\cal L}^\prime$ to conformal gravity. The reason is that, although ${\cal L}$ and ${\cal L}^\prime$ only differ by a total derivative, and therefore are equivalent as CFT's, this total derivative ceases to be a total derivative after replacing derivatives by covariant derivatives.
Hence, to distinguish between these two cases it is necessary to formulate the invariance directly at the level of the Lagrangian.

Under the variation \eqref{below} neither ${\cal L}$ nor  ${\cal L}^\prime$ are exactly invariant. Instead, they are both invariant up to a total derivative. Because ultimately we are interested in coupling these Lagrangians to conformal gravity, and anticipating on their future invariance under arbitrary coordinate transformations, we restrict the allowed total derivative and define an invariant Lagrangian as a Lagrangian whose variation is
\begin{align}\label{eq: Condition Inv Lag}
\delta \cal L = \partial_\mu \left(\xi^\mu \cal L\right)\,.
\end{align}
The rule is then that we can only couple those Lagrangians to conformal gravity that are invariant by themselves in the sense of \eqref{eq: Condition Inv Lag}. Note that in the case discussed above $\delta \cal L' \neq \partial_\mu \left(\xi^\mu \cal L'\right)$. In the remaining, when we talk about an invariant Lagrangian we will always mean invariance up to the total derivative \eqref{eq: Condition Inv Lag}.


As a last remark, let us stress that the relation between Poincar\'e invariants and scalar CFT's is not one-to-one. Although to each independent CFT we can associate a unique (linearly independent) Poincar\'e invariant, there exist Poincar\'e invariants that do not have a corresponding scalar CFT. Furthermore, not every CFT corresponding to a Poincar\'e-invariant has a kinetic term. The latter situation is what we denote as `potential terms' and the former  as `curvature terms'. Accordingly, in the remainder of this paper we will distinguish between three different categories of CFT's/spacetime invariants:
\begin{enumerate}


\item {\bf Potential terms} \\
Not every spacetime invariant corresponds to a CFT with a kinetic term. For instance, a cosmological constant term $\Lambda$ corresponds to a CFT without a kinetic term. Explicitly, we have
\begin{align}
\mathrm{CFT}_0: \;\; {\cal L} = \Lambda\,\phi^2\,,\quad w=-\tfrac{D}{2} \hskip 1truecm \Leftrightarrow \hskip 1truecm \mathrm{P}_0: \;\; e^{-1} {\cal L} = \Lambda \,,
\end{align}
where, for clarity, we set $\kappa=1$ from now on. Note that in this case the field redefinition \eqref{eq: field redef} is modified, as can be seen from the value of dilatation weight.


\item {\bf Kinetic terms} \\
This category includes all CFT's with time derivatives. An example is the CFT \eqref{scalarcft} corresponding to the Einstein-Hilbert invariant. One can also consider higher-derivative CFT's corresponding  to (linear combinations of) the invariants $R^{\mu\nu}R_{\mu\nu}$ and $R^2$ as will be discussed below.

\item {\bf Curvature terms} \\
There are spacetime invariants that do not correspond  to any CFT. For instance, starting from  the Weyl tensor squared in $D$ dimensions and making the redefinition \eqref{rescaling} one ends up with the following term:
\begin{align}
e^{-1} {\cal L} \sim \phi^{2\tfrac{D-4}{D-2}}\big(C_{\mu\nu}{}^{\text{\tiny AB}}\big)^2\,.
\end{align}
Clearly,  upon gauge-fixing $e_\mu{}^{\text{\tiny A}} = \delta_\mu{}^{\text{\tiny A}}$, this term vanishes and therefore does not lead to any non-trivial CFT.

\end{enumerate}





\subsection{Higher derivative invariants}

The procedure outlined above can easily be extended to include higher-derivative Poincar\'e invariants such as $R^2$ and $R^{\mu\nu}R_{\mu\nu}$.\,\footnote{The same does not apply to  the Riemann curvature squared term since that corresponds, using the terminology introduced above, to a curvature term.}
In that case one ends up with a higher-derivative scalar CFT.
For completeness, and in order to better illustrate some of the points just mentioned, we briefly discuss the Poincar\'e invariants corresponding to the CFTs with four derivatives. Assuming $D>4$,
we consider in this case a compensating scalar $\phi$ of dilatation weight
\begin{align}\label{rel weight HO}
w = -\tfrac{1}{2}(D-4)\,.
\end{align}
 Given this weight, it follows that the higher derivative field theory built out of the operator box squared, ${\cal L} \sim \phi\square^2\phi$, is a CFT. We call it CFT$_2$ below. Using the general procedure outlined above, it can be seen to correspond to the following combination of the Ricci tensor squared and the Ricci scalar squared:
\begin{align}\label{eq: Rmunusquare}
\mathrm{CFT}_2:  \;\; {\cal L} = -\tfrac{(D-2)^2}{D-4} \phi \square^2\phi \qquad \Leftrightarrow \qquad \mathrm{P}_2:  \;\;  e^{-1} {\cal L} = R^{\mu\nu} R_{\mu\nu} -\tfrac{D^3-\,4(D-2)^2}{16(D-1)^2}R^2 \,.
\end{align}
The fact that both the Ricci tensor squared  and the Ricci scalar squared are separately invariant indicates that there is another independent CFT at that order.
Starting from the Poincar\'e invariants $R^{\mu\nu} R_{\mu\nu}$ and $R^2$ separately the corresponding CFTs could be obtained by introducing the compensating scalar via a similar decomposition as we did  in \eqref{rescaling} for the Einstein-Hilbert term. Here, in analogy to what we will do in the non-relativistic case, we discuss a different way to obtain this result by starting  from the CFT  point of view.

First, note that up to a partial differentiation, the CFT considered in \eqref{eq: Rmunusquare} is equivalent to the Lagrangian ${\cal L}^\prime \sim (\square\phi)^2$. The variation of  Lagrangian ${\cal L}^\prime$ is proportional to $\square\phi$ which is the conformal scalar theory extensively discussed in section \ref{EHinv}, and obviously cannot be coupled to conformal gravity  given the weight \eqref{rel weight HO}. However, ${\cal L}^\prime$ can be made  invariant by adding compensating terms.\footnote{We thank Mehmet \"Ozkan for useful comments on this construction.} From this we learn that there exists another CFT, denoted below as CFT$_3$, which can be coupled to conformal gravity and, with appropriate normalisation, correspond to the Poincar\'e invariant $R^2$ after gauge-fixing:
\begin{align}\label{eq: CFT3}
\mathrm{CFT}_3: \;\; {\cal L} = 16\, \tfrac{(D-1)^2}{(D-4)^2}\left(\square\phi + \tfrac{2}{(D-4)}\frac{\partial_\mu\phi\partial^\mu\phi}{\phi} \right)^2 \qquad \Leftrightarrow \qquad \mathrm{P}_3: \;\; e^{-1} {\cal L} = R^2\,.
\end{align}
The Lagrangians \eqref{eq: Rmunusquare} and \eqref{eq: CFT3} describe two independent scalar CFTs. They are the only two invariant Lagrangians with four derivatives acting on the scalar $\phi$ of fixed dilatation weight \eqref{rel weight HO}.
It is possible to obtain them starting from the most general linear combination of all possible terms that can be written down with four derivatives that have the correct scaling behaviour and by requiring that under the rigid conformal transformations the Lagrangian transforms according to
\eqref{eq: Condition Inv Lag}.
 We thus find that, up to the curvature terms discussed above, a classification of all inequivalent CFTs  correspond to all possible Poincar\'e invariants.

This concludes our  review of the relativistic conformal construction. In the next section we will discuss what happens in the non-relativistic case.



\section{A non-relativistic conformal method}\label{nonrelcons}
In section \ref{relcons} we reviewed how the Einstein-Hilbert action arises from a CFT of a free scalar with a two-derivative kinetic term.
Like the Poincar\'e group which can be extended to the conformal group, the centrally extended Galilei group can be extended to the Schr\"odinger group.
This prepares the base for our non-relativistic conformal construction both of Lagrangians that are invariant under local Galilean symmetries as well as of  field equations that are covariant under these symmetries. The latter leads to the construction of the curved space  Newton-Cartan Gravity (NCG) equations of motion with torsion discussed in section \ref{NC} while the former leads to the construction of Galilean invariants in section \ref{HLgravity} and their physical realization as Ho\v rava-Lifshitz Gravity (HLG) discussed in section \ref{Relation to HLgravity}.


\subsection{Newton-Cartan variables}
In general, the gauging of the Galilean algebra \cite{Hartong:2015zia,Andringa:2010it} leads to a set of independent gauge fields which are given by a time-like vielbein $\tau_\mu$ and  a spatial vielbein $e_\mu{}^a$ --- with $a$ referring to the spatial local Galilean frame, $a=1, \cdots, d$ --- which obey the following transformation rules
\begin{subequations}
	\label{NCvariables}
\begin{align}
\delta \tau_\mu &= 0\,,\\ 
\delta e_\mu{}^a &= \Lambda^{ab}e_\mu{}^b + \Lambda^a\tau_\mu\,,
\end{align}
\end{subequations}
where $\Lambda^{ab}$ and $\Lambda^a$ are the parameters of a (local) spatial rotation and a Galilean boost, respectively.
Furthermore, both gauge fields transform as covariant vectors under general coordinate transformations with parameter $\xi^\mu$.
The inverses $\tau^\mu$ and  $e^\mu{}_a$ are defined by
\begin{alignat}{2}
\tau^\mu \tau_\mu &= 1\,, & \qquad \qquad  \tau^\mu e_\mu{}^a &= 0 \,, \nonumber \\
\tau_\mu e^\mu{}_a &= 0 \,, & \qquad \qquad e_\mu{}^a e^\mu{}_b &= \delta^a_b \,. \label{projective-inverse}
\end{alignat}
They obey the following transformation rules:
\begin{subequations}
	\begin{align}
	\delta \tau^\mu &= - \Lambda^a e^\mu{}_a\,,\\[.1truecm]
	\delta e^\mu{}_a &= \Lambda_{ab}e^\mu{}_b\,,
	\end{align}
\end{subequations}
and can be used to form the temporal and spatial projections of a given tensor $T_\mu$;
 \begin{align}
 T_0 \equiv \tau^\mu T_\mu\quad \mbox{and}\quad T_a\equiv e_a^\mu T_\mu \qquad \mbox{such that} \qquad T_\mu = T_0\tau_\mu + T_a e^a_\mu\,.
 \end{align}
We will extensively use this notation below.

The geometric realization of these variables is given in terms of the Newton-Cartan geometry which describes a non-relativistic spacetime.
A difference with the relativistic case is that one cannot define a metric for the full spacetime. Requiring Galilean invariance one can only define a metric
$\tau_{\mu\nu}$ in the time direction and a metric $h^{\mu\nu}$ in  the spatial directions separately:
\begin{align}\label{updown}
\tau_{\mu\nu} = \tau_\mu \tau_\nu\,,\hskip 2 truecm h^{\mu\nu} = e^\mu{}_a e^\nu{}_b\, \delta^{ab}\,.
\end{align}
To define a metric with upper indices in the time direction and a metric with lower indices in  the spatial directions that is invariant under Galilean boosts one needs a vector field that transforms under Galilean boosts with a shift \cite{Hartong:2015zia}: \begin{align}\label{acc}
\delta M_\mu = \Lambda_ae_\mu{}^a\,.
\end{align}
This vector field should be considered together with $\tau_\mu$ and $e_\mu{}^a$ to define the full Newton-Cartan geometry. In fact, using this vector field we can simply define the following boost invariant hatted variables;\begin{subequations}\label{hattedvielbein}
\begin{align}
\hat\tau^\mu&=\tau^\mu+ e^\mu{}^a M_\mu\,,\label{tauhat}\\
\hat e_\mu{}^a&= e_\mu{}^a-\tau_\mu M_a\,.\label{ehat}
\end{align}
\end{subequations}
Note that this basis preserves all the orthonormality conditions of \eqref{projective-inverse}.

The vector field $M_\mu$ can be promoted to a gauge field $m_\mu$ by a St\"uckelberg mechanism as we will see later on.
The gauge field $m_\mu$ is associated to the central charge transformation in the centrally extended Galilean algebra, i.e.~the Bargmann algebra \cite{Andringa:2010it}. There are several reasons to work with the Bargmann algebra rather than Galilean algebra:

\begin{description}
\item{1.} The Lagrangian of a non-relativistic particle is only invariant under Galilean boosts up to a total derivative. This leads to a centrally extended Galilean algebra.

\item{2.} A vector field is needed to solve for the connection fields of spatial rotations and Galilean boosts in terms of the other gauge fields \cite{Andringa:2010it}.

\item{3.} The vector field is needed to define a metric with upper indices in the time direction and a metric with lower indices in  the spatial directions as stated above in eq.~\eqref{hattedvielbein}.
\end{description}

\noindent In this section we do not work with the Bargmann algebra but with the Sch\"odinger algebra which is a minimal  extension of it that serves our purposes. Therefore, the  gauge  field  $m_\mu$ we will be using has a Schr\"odinger origin. More information about Schr\"odinger gravity and the gauging of the Schr\"odinger algebra can be found in Appendix \ref{App A}.

\subsection{Compensating scalar field}

In order to relate a Galilean  invariant to a Schr\"odinger invariant field theory we need to introduce  compensating fields. Since we add two extra symmetries, dilatations and central charge transformations, we introduce two  real scalars $\varphi$ and $\chi$ or, equivalently, a complex scalar $\Psi$:
\begin{align}
\Psi =\varphi\,e^{i\chi}\,.
\end{align}
Any Galilean invariant can then be made invariant under local dilatations and central charge transformations by replacing the Galilean vielbein fields, which from now on we give a superscript {\small G}, by Schr\"odinger vielbein fields, which we indicate with a superscript {\small Sch}, and by replacing the vector field $M_\mu$ by the Schr\"odinger gauge field
$m_\mu$ as follows:
\begin{subequations}
	\label{dc}
\begin{align}
(\tau_\mu )^{\text{\tiny G} }&= \varphi^{-\frac2w} (\tau_\mu)^{\text{\tiny Sch}}\,,\hskip 1truecm (e_\mu{}^a)^{\text{\tiny G}} = \varphi^{-\frac1w} (e_\mu{}^a)^{\text{\tiny Sch}}\,,	\label{tauvsedc}\\
 M_\mu &= m_\mu -\frac1\mass\partial_\mu\chi\,,\label{M.vs.m}
\end{align}
\end{subequations}
where $\mass$ is an arbitrary parameter which is inserted to adjust the mass dimension and the dilatation weight $w$ is fixed up to a field redefinition ambiguity which is removed once we fix the invariant theory for $\Psi$, see section \ref{HLgravity}. Here we use an arbitrary weight for the scalar field to  avoid further field redefinition and to harmonize the notation with the next sections. It is important to note that we have used different scalings for $\tau_\mu$ and $e_\mu{}^a$ in eq's. \eqref{tauvsedc} since we want to consider  the  case of Schr\"odinger gravity at $z=2$, see appendix \ref{App A}.

The two compensating scalars $\varphi$ and $\chi$, the Schr\"{o}dinger vielbein gauge fields $(\tau_\mu)^{\text{\tiny Sch}}$\,, $(e_\mu{}^a)^{\text{\tiny Sch}}$ and the Schr\"{o}dinger central charge gauge field $m_\mu$  transform under spatial rotations and Galilean boosts like the fields $(\tau_\mu)^{\text{ \tiny G}}\,, (e_\mu{}^a)^{\text{\tiny G}}$ and $M_\mu$.
Furthermore, they transform under dilatations, with parameter $\Lambda_{\text{D}}$,  and central charge transformations,
 with parameter $\sigma$, such that the left-hand-side of \eqref{dc} is invariant under these transformations:
\begin{subequations}
	\label{non-rel-trans}
 \begin{align}
 \delta\Psi &=\big( w\Lambda_{\text{D}}+ i\mass\sigma\big)\Psi\,,\label{eq: nonrel trans psi}\\[.1truecm]
  \delta (\tau_\mu)^{\text{\tiny Sch}} &= 2\Lambda_{\text{D}}(\tau_\mu)^{\text{\tiny Sch}}\,,\hskip 1truecm
 \delta (e_\mu{}^a)^{\text{\tiny Sch}} = \Lambda_{\text{D}}(e_\mu{}^a)^{\text{\tiny Sch}}\,, \\[.1truecm] \delta m_\mu &= \partial_\mu\sigma\,.
 \end{align}
 \end{subequations}

After substituting the decompositions \eqref{dc} back into the Galilean invariant  we end up with an action that describes the coupling of
a complex scalar $\Psi$ to Schr\"odinger gravity. To obtain the complex scalar field theory before coupling to Schr\"odinger gravity we impose the following gauge fixing conditions:
\begin{align}\label{gfcds}
(\tau_\mu)^{\text{\tiny Sch}} = \delta_{\mu0}\,,\hskip 1.5truecm (e_\mu{}^a)^{\text{\tiny Sch}} = \delta_\mu{}^a\,,\hskip 1.5truecm m_\mu=0\,,
\end{align}
after which  we do not distinguish between flat and curved indices anymore. These gauge-fixing conditions imply
the following constraint equations for the symmetry parameters:\begin{subequations}
\begin{align}
&\partial_\mu\xi^0 + 2\Lambda_{\text{D}}\delta_{\mu 0}=0\,,\\[.1truecm]
&\partial_\mu\xi^a + \Lambda^{ab} \delta_\mu{}^b+ \Lambda^a\delta_{\mu 0} + \Lambda_{\text{D}}\delta_\mu{}^a = 0\,,\\[.1truecm]
&\partial_\mu\sigma + \Lambda_a\delta_\mu{}^a=0\,,
\end{align}
\end{subequations}
which can be solved
in terms of the following rigid Schr\"odinger transformations:
\begin{subequations}
\label{solution1}
\begin{align}
\xi^0(t) &= a^0 -2\lambda_{\text{D}}t + \lambda_{\text{K}}t^2\,,  \\[.1truecm]
 \xi^c (t,{\bf x})&= a^c -\lambda^{cb}x^b -\lambda^c\,t -\lambda_{\text{D}}x^c +\lambda_{\text{K}}tx^c\,, \\[.1truecm]
\Lambda^{ab} &= \lambda^{ab}  \,,  \\[.1truecm]
 \Lambda^a({\bf x}) &= \lambda^a -\lambda_{\text{K}}x^a\,, \\[.1truecm]
\Lambda_{\text{D}}(t) &=\lambda_{\text{D}} -\lambda_{\text{K}}t\,,  \\[.1truecm]
 \Lambda_{\text{K}} &=\lambda_{\text{K}}\,, \\[.1truecm]
  \sigma({\bf x}) &=\sigma_0 - \lambda_a x^a + \tfrac{1}{2}\lambda_{\text{K}}x^2\,,\label{solution2}
\end{align}
\end{subequations}
where $a^0\,,a^c\,, \lambda^{ab}\,, \lambda^a\,, \lambda_{\text{D}}\,,\lambda_{\text{K}}$ and $\sigma_0$ are the (constant) parameters of
time translations, spatial translations, spatial rotations, Galilean boosts, dilatations, special conformal transformations and
central charge transformations, respectively.

The gauge fixing conditions \eqref{gfcds} imply that after substituting the decompositions \eqref{dc} back into the Galilean invariant
one can ignore all terms with the derivative acting on one of the gauge fields. Starting from a general Galilean invariant one thus obtains the  field theory of a complex scalar
 with dilatation weight $w$ and central charge weight $\mass$ that is invariant under the following rigid Schr\"odinger transformations:
\begin{align}
\delta\Psi = \xi^0\partial_0\Psi + \xi^a\partial_a\Psi + \big(w\Lambda_{\text{D}}+ i \mass \sigma\big)\Psi\,,\label{deltapsi}
\end{align}
with $\xi^0\,, \xi^a\,,\Lambda_{\text{D}}$ and $\sigma$ given in eq.~\eqref{solution1}.

\subsection{Schr\"odinger gauge fields}
One may also do the reverse, i.e.~derive the Galilean invariant that is dual to a given SFT of a complex scalar. For this purpose, it is convenient to
first introduce the gauge fields of Schr\"odinger gravity and their transformation rules as they follow from gauging the $z=2$ Schr\"odinger algebra. As explained in appendix \ref{App A}, on top of the independent gauge fields $\tau_\mu$, $e_\mu{}^a$ and $m_\mu$ introduced above  --- from now on we omit the superscript ``{\small Sch}''--- we introduce four new Schr\"odinger gauge fields:
\begin{itemize}
\item[]  $\omega_\mu{}^{ab}$ - gauge field of {\sl spatial  rotations},
\item[]  $\omega_\mu{}^a$ - gauge field of {\sl boosts},
\item[] $b_\mu$ - gauge field of {\sl dilatations}  and
\item[] $f_\mu$ - gauge field of {\sl special conformal transformations},
\end{itemize}
which are dependent and can be expressed in terms of the independent gauge fields $e^a_\mu\,,\tau_\mu$ and $m_\mu$. The time component of the dilatation gauge field is an exception and remains independent but can be set to zero due to its transformation rule.
The transformation rules for all Schr\"odinger gauge fields are presented in eqs.~\eqref{gaugetrafoszis1} and \eqref{gaugetrafoszis2}. The full details including the expressions for all curvatures etc.~can be found in \cite{Bergshoeff:2014uea}. In appendix \ref{App A} we reproduce in our conventions the results of \cite{Bergshoeff:2014uea} which are necessary for our work.

The spatial components $b_a=e_a{}^\mu b_\mu$ of the
dilatation gauge field are dependent and  will play the role of torsion terms. Instead, the time component $\tau^\mu b_\mu$ transforms as a shift under the special conformal transformations. Like in the relativistic case, this symmetry is equivalent to the property that in any Galilean invariant the time-component of the dilatation gauge field $b_0=\tau^\mu b_\mu$ drops out while  the spatial components $b_a = e_a{}^\mu b_\mu$, which are dependent,
i.e.~$b_a=b_a(e,\tau)$, remain as torsion terms \cite{Bergshoeff:2014uea}. In contrast, the fate of the central charge gauge field is rather different.
This gauge field remains independent and is invariant under the special conformal transformations. It transforms as a shift under Galilean boosts.

With the above information at hand it is  straightforward to couple a given SFT of a complex scalar to Schr\"odinger gravity. In most cases, one just needs to replace all derivatives by covariant derivatives. For the construction of these SFTs and their coupling to Schr\"odinger gravity, see appendix \ref{Sch SFT} and section \ref{sec: Coupled SFT}, respectively. The corresponding Galilean invariant is then obtained by imposing the gauge-fixing condition
\begin{align}\label{gfc4}
\Psi=1\,,
\end{align}
which fixes both the dilatations and the central charge transformations.

Note that before gauge-fixing  $m_\mu$ is the gauge field of central charge transformations while after gauge-fixing it is invariant under central charge transformations, i.e.~it is not
a gauge field anymore but an ordinary vector field. To distinguish between the two cases we will denote the vector field after gauge-fixing with  $M_\mu$.
For the convenience of the reader we summarize here the status of the dilatation gauge field $b_\mu$, the central charge gauge field $m_\mu$ and the vector field  $M_\mu$ before and after gauge-fixing:
\bigskip

\noindent {\sl before gauge-fixing:}\ \ $b_\mu$ and $m_\mu$ are the independent gauge fields of dilatations and central charge transformations, respectively. The time component $b_0$ of the  dilatation gauge field is the only field that transforms under special conformal transformations. The gauge  field $m_\mu$ before gauge-fixing is related to the  vector field $M_\mu$ according to \eqref{M.vs.m}.
\bigskip

\noindent{\sl after gauge-fixing}: The time-component $b_0$ drops out while the space-components $b_a$ which are dependent give rise to torsion. After gauge-fixing there is no difference between $m_\mu$ and $M_\mu$. They both are vector fields which  transform in the same way according to \eqref{acc}.
\vskip .4truecm

As in the relativistic case, there is a correspondence between  Galilean invariants and  SFTs of the compensating complex scalar $\Psi$ introduced above. In appendix \ref{Sch SFT} we classify all
independent SFTs with up to two time derivatives and four spatial derivatives, distinguishing between purely potential terms, see subsection \eqref{ap: Pot term}, and kinetic terms, see subsection  \eqref{ap: Kin term}. The lowest order SFT with one time derivative kinetic term is the Schr\"odinger action, see eq.~\eqref{Sch action}. In section \ref{HLgravity}, we will use these results as a starting point to construct, following the procedure developed here, the general Galilean invariants with the same order of derivatives. On top of this, we will also obtain what we refer to as the curvature terms which can not be obtained by SFTs. Note that, unlike in the relativistic case,  a general Galilean invariant can have {\sl less} derivatives than the corresponding SFT. This is a simple consequence of the last equation of \eqref{dc} and is signaled  by the presence of the vector field $M_\mu$ in the Galilean invariant.

\section{Newton-Cartan gravity}\label{NC}
In the same way that Einstein gravity may be derived from gauging the relativistic Poincar\'e algebra, the Newton-Cartan formulation of Newtonian gravity can be obtained from gauging the Bargmann algebra,
 i.e.~the centrally extended Galilean algebra
\cite{DePietri:1994je,Duval:1983pb,Andringa:2010it}.
In this section we will discuss how to define a consistent set of equations of motion describing Newton-Cartan Gravity (NCG) \cite{Cartan:1923zea,Cartan:1924yea} with torsion. We will consider only a special kind of torsion, called `twistless torsion', which in this context was first discussed in  \cite{Christensen:2013lma}.
The torsionless equations can be found in the original literature, see, e.g., \cite{Ehlers, Trautman}. To derive an extension of these equations of motion with torsion, it is very convenient to apply the non-relativistic conformal method developed in section \ref{nonrelcons}.
 We note that the torsionless equations of motion can be obtained by taking the non-relativistic limit of the Einstein equations \cite{Dautcourt,Bergshoeff:2015uaa}.\,\footnote{We thank Eric Spreen for a discussion on this point.}
 Since the Schr\"odinger algebra cannot be obtained as the contraction of a relativistic algebra it is not clear a priori whether the NCG equations of motion with torsion
  we will construct can similarly be obtained as the limit of some relativistic equation of motion.
 Since there is no NCG action available we will apply the non-relativistic conformal  method at the level of the equations of motion. In subsection \ref{sec:1} we will first explain the procedure by reproducing  the torsionless equations of motion. Then, in subsection \ref{sec:2} we will
 extend this result and construct the NCG equations of motion with twistless torsion.


\subsection{Torsionless NCG}\label{sec:1}
 In this subsection we first discuss the torsionless case.
 In the absence of torsion  the Galilean timelike vielbein field $(\tau_\mu)^{\text{\tiny G}}$  satisfies the  foliation constraint
\begin{equation}\label{constraint no torsion}
\partial_\mu (\tau_\nu)^{\text{\tiny G}} - \partial_\nu (\tau_\mu)^{\text{\tiny G}} =0\,,
\end{equation}
while the NC equations of motion are given by \cite{Ehlers,Trautman,Andringa:2010it}\begin{subequations}
	\label{NotorsionNCeom}
\begin{align}
(\tau^\mu)^{\text{\tiny G}} (e^\nu{}_a)^{\text{\tiny G}} \mathcal{R}_{\mu \nu}{}^a(G)&=0\,,\label{NCeom2}\\[.1truecm]
(e^\nu{}_c)^{\text{\tiny G}} \mathcal{R}_{\mu \nu}{}^{ca}(J) &= 0\,,\label{NCeom5}
\end{align}
\end{subequations}
where $\mathcal{R}(G)$ is the NC curvature of Galilean boosts, see eq. \eqref{Barg curvaturesG}, and $\mathcal{R}(J)$ is the NC curvature of spatial rotations,
see eq. \eqref{Barg curvaturesJ}.

Although there is no known action that gives the equations of motion \eqref{NotorsionNCeom},  we can apply the non-relativistic technique as explained in section \ref{nonrelcons} to
the constraint \eqref{constraint no torsion} and to the  equations of motion \eqref{NotorsionNCeom}. We first consider the constraint \eqref{constraint no torsion}. Upon substitution of the decomposition \eqref{dc}
in the constraint \eqref{constraint no torsion} we obtain
\begin{equation}
\partial_\mu (\tau_\nu)^{\text{\tiny Sch}} - \partial_\nu (\tau_\mu)^{\text{\tiny Sch}}  + 2w^{-1}\varphi^{-1}\big((\tau_\mu)^{\text{\tiny Sch}}\partial_\nu\varphi - (\tau_\nu)^{\text{\tiny Sch}}\partial_\mu\varphi\big)
=0\,,
\end{equation}
where $w$ is the dilatation weight of the compensating scalar $\varphi$.
Next, imposing the gauge-fixing condition \eqref{gfcds} in the above equation, we obtain the following constraint:
\begin{equation}\label{constraintvarphi}
\partial_a\varphi=0\,.
\end{equation}
This constraint is invariant under the rigid conformal transformations
\begin{equation}\label{transvarphi}
\delta\varphi = \xi^0\partial_0\varphi + \xi^a\partial_a\varphi + w\Lambda_{\text{D}}\varphi\,,
\end{equation}
with $\xi^0\,, \xi^a$ and $\Lambda_{\text{D}}$ given in eqs.~\eqref{solution1}. Making the same decomposition and imposing the same gauge conditions in the
equation of motion \eqref{NCeom5} leads to an expression that is proportional to the constraint \eqref{constraintvarphi} and hence is automatically satisfied.

We next apply the same manipulations to the remaining NC equation of motion \eqref{NCeom2}. After fixing the dilatation weight to $w=1$, this leads to the following equation for $\varphi$:\,\footnote{This equation was derived together with Jan Rosseel.}
\begin{equation}\label{2dconstraint}
\partial_0\partial_0\varphi=0\,.
\end{equation}
One can show that this equation, together with the constraint equation \eqref{constraintvarphi}, is invariant under the rigid conformal transformations
\eqref{transvarphi} for $w=1$. The two equations together define the SFT that underlies the torsionless NCG equations of motion.

It is straightforward  to recover the torsionless NC equation of motion \eqref{NotorsionNCeom}  from the SFT  defined by eqs.~\eqref{constraintvarphi} and \eqref{2dconstraint}. We first define the covariant derivatives, taken for $w=1$,
\begin{eqnarray}
D_0\varphi &=& \tau^\mu\big(\partial_\mu - wb_\mu\big)\varphi\,,\\[.1truecm]
D_a\varphi &=& e^\mu{}_a\big(\partial_\mu - wb_\mu\big)\varphi\,. \label{eq: def Davarphi}
\end{eqnarray}
Due to the presence of $b_0$ in the time covariant derivative which transforms as a shift under special conformal transformations, we obtain the following second order in time covariant derivative, taken for $w=1$,
\begin{equation}\label{confbox}
D_0D_0\varphi = \tau^\mu\bigg(\partial_\mu D_0\varphi -(w-2)b_\mu D_0\varphi + \omega_\mu{}^aD_a\varphi + w f_\mu\varphi\bigg)\,.
\end{equation}
A remark about the status of the time component $f_0=\tau^\mu f_\mu$ of the gauge field $f_\mu$ of special conformal transformations is in order here. When constructing Schr\"odinger gravity, this gauge field was dependent, see eq.~\eqref{eq:vfz=2}. In the case of zero torsion, i.e. $b_a=0$, the last two terms in the expression
for $f_0$ are proportional to the NCG equation of motion \eqref{NCeom5}.
Therefore, in this particular case there is no need to add these two terms to the definition of $f_0$. Instead, to obtain invariant equations of motion, it is sufficient to include \eqref{NCeom5} to the set of torsionless NCG equation of motion.

We thus end up with the following two equations describing the coupling of the compensating scalar $\varphi$ of dilatation weight $w=1$ to Schr\"odinger gravity:
\begin{equation}\label{vonfeom}
D_0D_0\varphi =0\,,\hskip 1.5truecm D_a\varphi =0\,,
\end{equation}
with $f_0$ defined by eq.~\eqref{eq:vfz=2} with the assumption that  eq.~\eqref{NCeom5} is one of the NCG equations of motion. Note that in the torsionless case the two
Schr\"odinger curvature terms in eq.~\eqref{eq:vfz=2} effectively reduce to two Galilean curvature terms since $b_a=0$ and $b_0$ drops out from the final equations of motion due to the invariance under special conformal transformations. Therefore, on-shell we can set the two curvature terms in eq.~\eqref{eq:vfz=2} equal to zero.
It can be checked that the two equations \eqref{vonfeom} together are invariant under all local Schr\"odinger transformations for $w=1$. Imposing the gauge-fixing condition
\begin{equation}\label{gfc2}
\varphi=1\,,
\end{equation}
and using the explicit expressions \eqref{dependentgaugefields}-\eqref{eq:vfz=2} together with eq.~\eqref{NCeom5} the two equations \eqref{vonfeom} reduce to
\begin{equation}\label{final1}
 \tau^\mu e^\nu{}_a \mathcal{R}_{\mu \nu}{}^a(G)=0\,,\hskip 1.5truecm b_a=0\,,
\end{equation}
which indeed is the NC equation of motion \eqref{NCeom2} together with the statement of zero torsion. Along with our assumption that eq.~\eqref{NCeom5} is valid
we thus have reproduced the full set of torsionless NCG equations of motion.

\subsection{Torsional NCG}\label{sec:2}
 In this subsection  we consider the general case in which  there is nonzero torsion, i.e.~$b_a\ne 0$.  This case is more complicated due to the fact that we a priori neither know the torsional NCG equations of motion nor the corresponding SFT. Of course, we require that the final result, taken for $b_a=0$, reduces to the torsionless NCG equations of motion  given in the previous subsection. It turns out that the torsion extension of the NCG equation \eqref{NCeom5} does not lead to any SFT equation in the same way that a curvature invariant does not correspond to a SFT action.\,\footnote{If one would try to derive a SFT equation corresponding to the NCG equation \eqref{NCeom5} one would fail due to the fact the the leading $\mathcal{R}_{\mu a}(J)$ term decouples from the SFT equation. In the case of a space/time projection of this equation, this leads to a problem since the corresponding curvatures $\mathcal{R}_{0a}(J)$ is not invariant by it-self.}


To derive the extension with torsion, it  is easiest to approach the issue from the SFT side. One thing that changes in the torsional case  is that the vielbein now satisfies a dilatation-covariant foliation constraint expressing the fact that the torsion is twistless:
\begin{align}\label{constraint3}
\partial_\mu (\tau_\nu)^{\text{\tiny G}} - \partial_\nu (\tau_\mu)^{\text{\tiny G}} -2b_a(e_{\mu}{}^a)^{\text{\tiny G}}(\tau_\nu)^{\text{\tiny G}}
 +2b_a(e_\nu{}^a)^{\text{\tiny G}}(\tau_\mu)^{\text{\tiny G}}=0\,.
\end{align}
This constraint is enough to guarantee that the foliation space is a Riemannian manifold.
Making a decomposition and gauge-fixing does not lead to the constraint \eqref{constraintvarphi} anymore as the result of this procedure on \eqref{constraint3} is automatically vanishing.
In the absence of this constraint the equation of motions are no longer invariant under the Galilean symmetries. Equivalently, from the scalar field theory point of view, eq.~\eqref{2dconstraint} is not invariant under rigid conformal transformations anymore:
\begin{align}
\delta \big(\partial_0\partial_0\varphi\big ) = -2\lambda^a\partial_0\partial_a\varphi\,.
\end{align}

To make equation \eqref{2dconstraint} invariant under all rigid Schr\"odinger transformations, we introduce the second compensating scalar $\chi$ which transforms under rigid conformal transformations as follows:
\begin{align}\label{transchi}
\delta\chi = \xi^0\partial_0\chi + \xi^a\partial_a\chi + \mass\sigma\,,
\end{align}
with $\xi^0\,, \xi^a$ and $\sigma$ given in eq.~\eqref{solution1} and where $\mass$ is arbitrary. Under Galilean boosts the spatial derivative of $\chi$ transforms with an inhomogeneous term
\begin{align}
\delta_{\lambda^a}\partial_a \chi \sim  -\mass\lambda_a.
\end{align}
This means that at lowest order eq.~\eqref{2dconstraint}  can  be made invariant under Galilean boosts by adding a $\chi$-term to that equation.
Pursuing this iterative procedure we find the following  Schr\"odinger invariant field equation:
\begin{align}
\partial_0\partial_0\varphi -\frac{2}{\mass} (\partial_0\partial_a\varphi )\partial_a\chi +\frac{1}{\mass^2}  (\partial_a\partial_b\varphi)\partial_a\chi\partial_b\chi&=0\,.
\label{constraintrigid1}
\end{align}
This is the  SFT equation that underlies the torsion extension of the  NCG equation of motion \eqref{NCeom2}.

To extract the explicit form of this torsion extension, we promote the rigid Schr\"odinger symmetry  of the SFT equation \eqref{constraintrigid1} to a local one by coupling
the two compensating scalars to Schr\"odinger gravity. This can be achieved by replacing all ordinary derivatives by Schr\"odinger-covariant ones:\footnote{Note that all covariant derivatives commute such that there is no ambiguity in this procedure, see eq.~\eqref{eq: Covcommute}.}
\begin{align}
D_0D_0\varphi -\frac{2}{\mass} (D_0D_a\varphi )D_a\chi +\frac{1}{\mass^2}  (D_aD_b\varphi)D_a\chi D_b\chi&=0\,.
\label{constraintlocal1}
\end{align}
We have used here the following definitions of Schr\"odinger-covariant derivatives (taken for $w=1$):
\begin{eqnarray}
D_0D_a\varphi &=& \tau^{\mu}\big(\partial_{\mu}D_a\varphi - (w-1)b_{\mu}D_{a}{\varphi} - \omega_{\mu a}{}^{b}D_b\varphi\big)\,,\\
D_aD_b\varphi &=& e_{a}^{\mu}\big(\partial_{\mu}{D_{b}{\varphi}} - (w-1)b_{\mu}D_{b}{\varphi} - \omega_{\mu b}{}^{c}D_c{\varphi}\big) \,,\label{eq: Def DaDbvarphi}\\
D_a\chi &=&  e_{a}^{\mu}\big(\partial_{\mu}{\chi} - \mass m_{\mu}\big)\,.
\end{eqnarray}

Imposing the gauge-fixing conditions
\begin{align}
\varphi=1\,,\hskip 2truecm \chi=0
\end{align}
in eq.~\eqref{constraintlocal1} we obtain the following torsion extension of the NCG equation of motion \eqref{NCeom2}:
\begin{eqnarray}
\tau^\mu\Big(\mathcal{R}_{\mu a}{}^a (G) + 2M^b\mathcal{R}_{\mu a}{}^a{}_b(J)\Big)+  M^bM^c\mathcal{R}_{ba}{}^a{}_c(J)- 2M^a K_a +  \mathcal{D}_a b_a M^bM^b \nonumber\\ \qquad\qquad + \;2\Omega^a_\mu \big( - \tau^\mu  b_a + b_b e_b^\mu  M^a - e_a^\mu b_b M^b\big) = 0\,,\qquad \label{NCeom4}
\end{eqnarray}
where $\mathcal{D}$ is the Galilean covariant derivative, see the definition \eqref{def. Galcovder1}, and $K_a$ is defined in eq. \eqref{def. Galinv1} as the Galilean boost invariant version of $\mathcal{D}_0b_a$.

To obtain the torsional extension of the NCG equation of motion \eqref{NCeom5} we replace the Galilei curvature $\mathcal{R}(J)$ by the corresponding Schr\"odinger curvature $R(J)$, see eqs.~\eqref{Omegas1} and \eqref{SchcurvatureRJ},
\begin{eqnarray}
e^\nu{}_c R_{\mu \nu}{}^{cb}(J) &=& 0\,.\label{NCeom6}
\end{eqnarray}
Since this equation transforms covariantly under dilatations, it does not lead to a corresponding SFT equation.
 Note that, upon setting the torsion equal to zero, i.e.~$b_a=0$, the equations of motion \eqref{NCeom4}  and \eqref{NCeom6} reduce to the torsionless equations of motion  given in eqs.~  \eqref{final1} and \eqref{NCeom5} respectively.

A drawback of the equations of motion  \eqref{NCeom4} and \eqref{NCeom6} is that the Galilean invariance is not manifest. One can show that the  expression \eqref{NCeom4} can be rewritten in the following  manifestly Galilean invariant form:
\vskip -.1truecm
\begin{empheq}[box=\widefbox]{equation}
\hat\tau^\mu \partial_\mu K + K^{ab}K_{ab} - \triangle \Phi - 8\,\Phi\, b\cdot b - 2\,\Phi\, \mathcal{D}\cdot b - 6\,b^a\mathcal{D}_a\Phi = 0\,. \label{eq: NCeom Gal inv}
\end{empheq}
\vskip .2truecm
\noindent The dot in equation \eqref{eq: NCeom Gal inv} refers to the contraction of the spatial indices, i.e. $b\cdot b\equiv\delta^{ab}b_ab_b$ and $\mathcal{D}\cdot b\equiv\delta^{ab}\mathcal{D}_ab_b$. We recall that $\hat \tau^\mu$ is the Galilean invariant defined by eq.~\eqref{tauhat}. The field $\Phi$ is given by
\begin{equation}\label{eq: def Phi}
\Phi \equiv M_0 + \frac{1}{2}\delta^{ab}M_a M_b\,.
\end{equation}
After gauge-fixing to a frame with constant acceleration, $\Phi$ can be identified with the Newton potential. The Galilean covariant derivatives of $\Phi$ are given by
\begin{align}
\mathcal{D}_a\Phi = e^\mu_a \partial_\mu \Phi\,,\qquad
\triangle\Phi \equiv \delta^{ab}\mathcal{D}_a\mathcal{D}_b\Phi = e^{\mu a}\partial_\mu \mathcal{D}_a\Phi -  e_a^\mu\Omega^{ab}_{\mu}\mathcal{D}_b\Phi \,,
\end{align}
while $K \equiv \delta^{ab}K_{ab}$ and $K_{ab}$ is defined in eq. \eqref{def. Galinv2} as the Galilean boost invariant version of $\mathcal{D}_aM_b$, a definition that we repeat here for the convenience of the reader
\begin{align}
K_{ab}&=\cD_aM_b+M_ab_b+M_bb_a\,.\label{eq: hat Omegaab}
\end{align}
Here $\mathcal{D}_aM_b$ is the Galilean covariant derivative defined in eq.~\eqref{def. Galcovder2}. Note that $K_{ab}$ is symmetric as a consequence of the symmetric nature of $\mathcal{D}_aM_b$, as shown in eq.~\eqref{eq: antisym part DM}, and it can be seen as an extrinsic curvature. One may verify that all the terms appearing in \eqref{eq: NCeom Gal inv} are invariant by themselves. The same can be done for the remaining equations of motion \eqref{NCeom6}. The two different projections of this equation can be  rewritten as
\vskip -.2truecm
\begin{subequations}\label{torsionalNCGRJ}
\begin{empheq}[box=\widefbox]{align}
\mathcal{D}^bK_{ab} -\mathcal{D}_aK + K_{ab}b^b -b_aK+(d-1)K_a &= 0\,, \label{eq: NCeom Gal inv2} \\[.2truecm]
\mathcal{R}_{ac}{}^{c}{}_b(J) + (d-2)(\mathcal{D}_ab_b + b_ab_b)+ \delta_{ab}\left(\mathcal{D}\cdot b -(d-2)b\cdot b\right) &= 0\,,\label{eq: NCeom Gal inv3}
\end{empheq}
\end{subequations}
\vskip .2truecm
\noindent where \eqref{eq: NCeom Gal inv2} has been obtained by contracting \eqref{NCeom6} with $\hat \tau^\mu$ and \eqref{eq: NCeom Gal inv3} is obtained by contracting \eqref{NCeom6} with $e_a^\mu$.\footnote{The contraction by $\hat \tau^\mu$ is necessary in order to make \eqref{eq: NCeom Gal inv2} manifestly Galilean invariant by itself.} The Galilean covariant derivative of $K_{ab}$ is given by
\begin{align}
\mathcal{D}_a K_{bc} = e_a^\mu\left(\partial_\mu K_{bc} - \Omega_{\mu}{}_b{}^dK_{d c} - \Omega_\mu{}_c{}^{d}K_{b d} \right)\,.
\end{align}
In order to show that \eqref{NCeom5} is recovered from eqs.~\eqref{torsionalNCGRJ} in the torsionless limit $b_a=0$
one needs to use the identity:
\begin{align}
\hat \tau^\mu \mathcal{R}_{\mu c}{}^{c}{}_a(J) & = \mathcal{D}^bK_{ab} -\mathcal{D}_aK + K_{ab}b^b -b_aK + M^b\mathcal{D}_{b}{b}_{a} \nonumber\\[.2truecm]
& \quad - M_a(\mathcal{D}\cdot b + b \cdot b) + {b}_{a} {b}_{b} M^b + \Omega^{b}_{\mu}e_{b}^{\mu}b_{a} - \delta_{a b}\Omega^{b}_{\mu}e_{c}^{\mu}b^c\,.\quad \label{eq: Identity Cov K}
\end{align}
The equations \eqref{eq: NCeom Gal inv} and \eqref{torsionalNCGRJ} are our final result for the torsional NCG equations of motion. To derive these equations, it was very convenient to have the underlying SFT in mind. Note that equation \eqref{eq: NCeom Gal inv} consists of six terms. The term $\triangle\Phi$ yields, after gauge-fixing to an earth-based frame, the Poisson equation of the Newton potential to leading order. Two terms are proportional to the extrinsic curvature and the remaining three terms are all proportional to torsion. Note also that the equation \eqref{eq: NCeom Gal inv2} is invariant under time-reversal symmetry since both $K_a$ and $K_{ab}$ are odd under that symmetry. As far as we know this general equation has not appeared before in the literature. It would be very interesting to find a non-relativistic situation where these equations of motion should be used, for instance in condensed matter systems,  and to construct solutions to these equations.

\section{Ho\v rava-Lifshitz gravity}\label{HLgravity}
In this section we apply the non-relativistic conformal formalism developed  in section \ref{nonrelcons} to SFTs of a compensating complex scalar  to obtain a number of (higher-derivative) Galilean invariant actions.
To do so, we will first in subsection \ref{sec: Coupled SFT} couple the scalar SFTs of appendix \ref{Sch SFT} to Schr\"odinger gravity. In the next subsection \ref{gaugefixing}, we will gauge-fix the dilatation and central charge transformations in order to obtain various Galilean invariants.
Next, in section \ref{Relation to HLgravity} we will make contact between the  higher-derivative Galilean invariants we construct and $z=2$  Ho\v rava-Lifshitz gravity. Our results will be in  agreement with those of \cite{Hartong:2015zia}.

\subsection{Scalar coupled Schr\"odinger gravities}\label{sec: Coupled SFT}
The aim of this subsection is to classify all possible scalar actions that are invariant under the local Schr\"odinger transformations. To obtain this result, we start with the SFTs  classified in appendix \ref{Sch SFT} and  couple  them to Schr\"odinger gravity, whose construction is summarized in appendix \ref{App A}. For presentation purposes this procedure is divided into the case of actions with and without time derivatives. We start with the simpler case of purely spatial derivatives in subsubsection
\ref{sec: Coupled Pot}. Next, we  consider first and second order time-derivatives in subsubsection \ref{sec: Coupled Kin}. In a third subsubsection  \ref{sec: Coupled Cuv} we construct locally Schr\"odinger invariant curvature invariants that do not correspond to a SFT.

The coupling of the SFTs to Schr\"odinger gravity  is obtained by replacing the flat space derivatives $\partial_0$ and $\partial_a$ by the covariant derivatives $D_0=\tau^\mu (\partial_\mu + \dots)$ and $D_a=e_a^\mu (\partial_\mu+ \dots)$ and multiplying the flat space Lagrangian by the determinant $e=\mbox{det}(\tau_\mu,e^a_\mu)$. The inverse vielbeins represent the coupling to the Newton-Cartan background and the dots represent the set of gauge fields that need to be added for covariance. This procedure can be ambiguous for two distinct reasons. First, because the commutation properties of partial derivatives is in general lost for the covariant derivatives. It turns out that in most cases, at the order at which we are working, the covariant derivatives do commute and this ambiguity does not appear. However, this will not always be the case, see e.g.~the discussion around eq.~\eqref{eq: HighCovcommute}. The second source of ambiguity is related to the fact that the  SFT Lagrangians are defined only up to a total derivative. In order to deal with this ambiguity, we have to make sure that the Lagrangian itself is actually an invariant before we can proceed with covariantizing the SFT. Concretely, as explained in section \ref{relcons} and appendix \ref{Sch SFT}, respectively in the relativistic and non-relativistic cases, this is achieved by imposing the condition \eqref{eq: ap var L condition} to the variation of the SFT Lagrangians.

In order to remove possible field redefinitions, we will find it convenient to fix the dilatation weight of the complex scalar field $\Psi$ (and hence also of the real scalar $\varphi$) to
\begin{align}\label{eq: nonrel weight}
w = - \frac{d+2-2n_t-n_s}{2}\,,
\end{align}
where $n_t$ is the number of time derivatives and $n_s$ the number of spatial derivatives in a given term at a given order. We refer the reader to appendix \ref{Sch SFT} for more details.

\subsubsection{The potential terms}\label{sec: Coupled Pot}
In this section we collect the locally Schr\"odinger invariant scalar Lagrangians that are zeroth, second and fourth order in spatial derivatives. As follows from the analysis done in appendix \ref{Sch SFT}, the potential terms correspond to all the possible inequivalent ways we can act with spatial derivatives on the norm of the scalar field while all indices are contracted.

At the lowest order, with $n_t=0$ and $n_s=0$, the coupling of \eqref{eq: flat P0} only amounts to a multiplication by the determinant $e=\mbox{det}(\tau_\mu,e^a_\mu)$. Hence, we directly obtain
\begin{align}\label{Action P0}
S^{\tiny \text{(0)}}&= \int dtd^d{\bf x}\, e\, \Lambda_0 \varphi^2\,,
\end{align}
with $w=-\tfrac{d+2}{2}$ and $\Lambda_0$ an arbitrary constant.

With $n_t=0$, $n_s=2$ there are only two possible Lagrangians that lead to locally invariant actions. With the dilatation weight fixed to $w=-\frac{d}{2}$ according to \eqref{eq: nonrel weight}, these are $eD_a\varphi D_a\varphi$ and $e\varphi \triangle\varphi$ where we denote $\triangle\equiv \delta^{ab}D_a D_b$. The Schr\"odinger covariant derivatives are naturally defined from the transformation of the complex field $\Psi=\varphi e^{i\chi}$, see eq.~\eqref{deltapsi}, and have been given in eqs. \eqref{eq: def Davarphi} and \eqref{eq: Def DaDbvarphi}.
After coupling to Schr\"odinger gravity these two Lagrangians differ only by a boundary term:
\begin{align}\label{eq: 2nd order buliding Pot}
\varphi \triangle\varphi = - D_a\varphi D_a\varphi + e^{-1}\partial_\mu\left( e \varphi e^{a\mu}D_a\varphi\right)\,.
\end{align}
Hence, we can write a single locally invariant scalar potential action at this order,
\begin{align}\label{Action P1}
S^{\tiny \text{(1)}}&= \int dtd^d{\bf x}\, e\,D_a \varphi D_a \varphi \,,  
\end{align}
which is the coupled version of \eqref{eq: flat P1}.

With $n_t=0,n_s=4$ we can construct invariants by combining $D_a\varphi D_a\varphi$ and $\varphi \triangle\varphi$ given that in this case the boundary term of \eqref{eq: 2nd order buliding Pot} will have a non-trivial effect. Equivalently, this corresponds to coupling the invariant terms \eqref{eq: flat Ps} to Schr\"odinger gravity. Explicitly, we obtain\begin{subequations}\label{ActionPPP}
\begin{align}
S^{(2)} &= \int dtd^d{\bf x}\, e\,\varphi^{-2}\big(D_a \varphi D_a \varphi\big)^2\,, \label{Action P2} \\ 
S^{(3)}&= \int dtd^d{\bf x}\, e\,\varphi^{-1}\big(D_a \varphi D_a \varphi\big)\triangle\varphi\,, \label{Action P3} \\ 
S^{(4)}&= \int dtd^d{\bf x}\, e\,\big(\triangle\varphi\big)^2\,,  \label{Action P4}
\end{align}
\end{subequations}
where local invariance is achieved with the dilatation weight of $\varphi$ fixed to be $w=-\frac{d-2}{2}$. Note that a Lagrangian of the form $e\varphi \triangle^2 \varphi$ would only differ from the Lagrangian used in \eqref{Action P4} by a boundary term and does not yield a new independent invariant. Also, let us stress that, using the commutation properties of partial derivatives and performing some partial integrations, we could have rewritten the invariant \eqref{eq: flat P4} in such a way that after coupling to Schr\"odinger gravity we would obtain a different invariant;
\begin{align}\label{eq: S4prim}
S^{(4')}=\int dtd^d{\bf x}\, e\, D^a D^b\varphi \, D_aD_b \varphi\,.
\end{align}
However, due to the relation,
\begin{align}
D^{a}{D^{b}{\varphi}}D_a{D_b{\varphi}} =
(\triangle \varphi)^2
+ D_{a}{\varphi}D_{b}{\varphi}R^{ab}(J)
+ e^{-1}\partial_{\mu}\big[ee_{a}^{\mu}\left(D_b\varphi D_aD_b\varphi - D_a\varphi\triangle\varphi\right)\big]\,,
\end{align}
the action \eqref{eq: S4prim} only differs from the invariant \eqref{Action P4} by a curvature term. We will classify such curvature terms later in subsubsection \ref{sec: Coupled Cuv}.

Strictly speaking, these are not all possible ways to form independent locally Schr\"odinger invariant scalar field theories with $n_t=0,n_s=4$. We could also construct potential terms with the spatial derivative acting on $\chi$. This analysis is done in appendix \ref{ap: Pot term} and leads to an extra action, the SFT given in \eqref{eq: flat P5}. Although this Lagrangian is invariant in the sense of \eqref{eq: Condition Inv Lag} and hence can straightforwardly be coupled to Schr\"odinger gravity, we show in appendix \ref{ap: Kin term} that it no longer leads to an independent theory once we consider the SFT's with kinetic terms made of the complex field $\Psi$.

\subsubsection{The kinetic terms}\label{sec: Coupled Kin}
We now proceed to the coupling of the complex scalar field theories with exactly one and two time-derivatives. At first order in time derivatives the Schr\"odinger action \eqref{Sch action}, which we repeat here for clarity,
\begin{align}\label{Saction}
\mathrm{SFT}_5: \;\; S^{(5)} = \int dtd^d{\bf x}\, \Psi^\star\Big(i \partial_0  - \tfrac{1}{2{\mass}}\partial_a\partial_a \Big)\Psi\,,
\end{align}
describes, up to potential terms, the unique scalar field theory with $n_t=1$ that is Schr\"odinger invariant. The action \eqref{Saction} is invariant under the rigid Schr\"odinger symmetries \eqref{solution1}
for a complex scalar $\Psi$ of dilatation weight $w=-\tfrac{d}{2}$ and arbitrary central charge of weight $\mass$.

After coupling the Schr\"odinger action \eqref{Saction} to Schr\"odinger gravity we obtain
\begin{align}\label{Action S1}
S^{(5)} = \int dtd^d{\bf x}\, e\,\Psi^\star\Schbox \Psi\,,
\end{align}
with $w=-\tfrac{d}{2}$ and the Schr\"odinger covariant derivatives given by\begin{subequations}\label{covdiffsch1}
\begin{align}
\Schbox \Psi &\equiv\left(i D_0  - \tfrac{1}{2{\mass}} \triangle\right) \Psi \,,\label{eq: def Schbox}\\[.1truecm]
D_0\Psi &=  \tau^\mu\big(\partial_\mu -w  b_\mu -i{\mass}m_\mu\big)\Psi\,, \\[.1truecm]
D_{a}\Psi &= e^{\mu}{}_{a}\big(\partial_{\mu}-wb_{\mu}-i{\mass}m_{\mu}\big)\Psi\,,\\[.1truecm]
\triangle\Psi &= e^{\mu}{}_{a}\Big[(\partial_{\mu}-( w-1)b_{\mu}-i{\mass}m_{\mu})D_{a} -\omega_{\mu a}{}^{b}D_{b}+i{\mass}\omega_{\mu a}\Big]\Psi\,.
\end{align}
\end{subequations}
The local invariance of \eqref{Action S1} can be easily checked using the transformation rule
\begin{align}
\delta \left(\Psi^\star\Schbox\Psi\right)&=2({w}-1)\Lambda_D \Psi^\star \Schbox\Psi -i\left({w}+\tfrac{{d}}{2}\right)\Lambda_K \Psi^\star \Psi\,. \label{eq: variation BoxSch}
\end{align}
Using the fact that the weight of the determinant $e=\mbox{det}(\tau_\mu,e^a_\mu)$ is $d+2$, the action  \eqref{Action S1} is indeed invariant under local Schr\"odinger transformations for $w=-\tfrac{d}{2}$. Note that although the action \eqref{Action S1} is not manifestly real its imaginary part is a boundary term.

We will see that the Galilean invariant corresponding to the Schr\"odinger action will have inconsistent equations of motion by itself. However, this invariant can be added to  the  Galilean invariants
with two time-derivatives that we will construct below.\,\footnote{This is similar to the cosmological constant term that, by itself, has an inconsistent equation of motion, but nevertheless can be added to the Einstein-Hilbert action.\label{ftn}} Such higher-order Galilean invariants are needed in order to reproduce the kinetic terms of  $z=2$ Ho\v rava-Lifshitz gravity, see the actions \eqref{Action S2} and \eqref{Action S3}.

We next consider the Schr\"odinger scalar theories with two time derivatives, i.e.~$n_t=2$. To be concrete, we consider the three scalar SFT's classified in appendix \ref{ap: Kin term}, see eq's.~\eqref{Sch action 2}, \eqref{Sch action 3} and \eqref{Sch action 4} and couple them to Schr\"odinger gravity. We thus obtain the following three locally Schr\"odinger invariant actions for a complex scalar field of dilatation weight $w=-\tfrac{d-2}{2}$ and arbitrary central charge weight $\mass$
\begin{subequations}
\label{2ndorderschrgr}:	
\begin{align}
S^{(6)} &= \int dtd^d{\bf x}\, e\,\Psi^\star \Schbox^2 \Psi\,,  \label{Action S2}\\
S^{(7)}&=\int dtd^d{\bf x}\, e\, \Big| \Schbox \Psi  + \frac{1}{\mass d}\Big(\triangle \Psi - \frac{D_a \Psi D_a \Psi}{\Psi} \Big) \Big|^2\,, \label{Action S3}\\
S^{(8)}&=\int dtd^d{\bf x}\, e\,  (\Psi^\star\Psi)^{-1}\Big( i\Psi^\star D_0 \Psi - i\Psi D_0 \Psi^\star +  \frac{D_a \Psi D_a \Psi^\star}{\mass} \Big)^2 \,, \label{Action S4}
\end{align}
\end{subequations}
where  $|\cdot|$ denotes the norm, e.g.~$|\Psi|^2=\Psi^\star\Psi$.
There are no ambiguities related to the order of the derivatives in the process of replacing the partial derivatives of the Lagrangians \eqref{Sch action 3} and \eqref{Sch action 4} by the covariant derivatives of \eqref{Action S3} and \eqref{Action S4} due to the identities
\begin{align}
[D_a, D_b]\Psi = 0\,, \qquad \qquad [D_0,D_a]\Psi = 0\,.\label{eq: Covcommute}
\end{align}
However, for the action \eqref{Action S2} the order of the covariant derivatives does matter, as can be seen from the non-vanishing of the higher order commutation relation
\begin{align}
[D_0,\triangle]\Psi = R(J)_{0b}{}^{ab}(D_a +2i\mass M_a)\Psi+ i\mass R(J)_{cb}{}^{ab}M_aM^c\Psi \label{eq: HighCovcommute}\,,
\end{align}
where $R(J)_{\mu\nu}{}^{ab}$ is the Schr\"odinger spatial rotation curvature defined in eq.~\eqref{Schcurvature}. Hence, in this particular case, it is the invariance under local Schr\"odinger transformations that ultimately fixes the correct order in which the covariant derivatives need to appear. We confirm below that, unsurprisingly, $\Schbox^2$ turns out to be the correct combination.

The higher order covariant derivatives are constructed by first determining the transformation rules of the lower-order covariant derivatives acting on $\Psi$.
In particular,
the higher order covariant derivatives acting on $\Psi$ that occur in eq.~\eqref{Action S2} are given by\begin{subequations}
\begin{align}
D_0^2\Psi&=\tau^\mu\Big[(\pa_\mu-( w-2)b_\mu-i\mass\, m_\mu)D_0 +\omega_\mu{}^aD_a + w f_\mu\Big]\Psi\,,\\[.1truecm]
D_0 \triangle\Psi&=\tau^\mu\Big[ (\pa_\mu-( w-2)b_\mu-i\mass\, m_\mu)\triangle +i\mass\left(2\omega_\mu{}^aD_a - d \,f_\mu\right)\Big]\Psi\,,\\[.1truecm]
\triangle D_0\Psi&=e^\mu{}_a\Big[ (\pa_\mu-( w-3)b_\mu-i\mass\, m_\mu)D_a D_0 -\omega_\mu{}_a{}^bD_b D_0 +\omega_\mu{}^bD_a D_b\nn
&\qquad\qquad\qquad\qquad+( w-1)f_\mu D_a+i\mass\,\omega_\mu{}_aD_0\Big]\Psi\,,\\[.1truecm]
\triangle^2\Psi&=e^\mu{}_a\Big[ (\pa_\mu-( w-3)b_\mu-i\mass\, m_\mu)D_a \triangle-\omega_\mu{}_{ab}D^b\triangle\nn
&\qquad\qquad\qquad\qquad+i\mass\left(\omega_\mu{}_a\triangle+2\omega_\mu{}^bD_aD_b-( d+2)f_\mu D_a\right)\Big]\Psi \,.
\end{align}
\end{subequations}

A few remarks are in order. The invariance of the action \eqref{Action S2} under Schr\"odinger symmetries can be confirmed from the transformation rule
\begin{align}\label{eq: delta schbox^2}
\delta (\Psi^\star\Schbox^2\Psi )=({w}-4)\Lambda_D\Psi^\star\Schbox^2\Psi-i\left(2{w}-2+{d}\right)\Lambda_K\Psi^\star\Schbox\Psi\,,
\end{align}
where the last term of \eqref{eq: delta schbox^2} drops given that the dilatation weight is fixed to $w=-\tfrac{d-2}{2}$.

It is  instructive to consider the variation of the second order covariant time derivative
\begin{align}
\delta \left(D_0^2\Psi\right) =\big[({w}-4)\Lambda_D+i{\mass}\sigma\big]D_0^2\Psi-2\lambda^aD_aD_0\Psi-2( w-1)\Lambda_KD_0\Psi\,.\label{D02}
\end{align}
As we confirmed, the action \eqref{Action S2} contains precisely the terms needed to compensate for the variation \eqref{D02}.
Nevertheless, we note that the last term in \eqref{D02} could also vanish with a dilatation weight $w=1$. This fact has actually been used in section \ref{NC} in the context of the invariance of the Newton-Cartan equations of motion which has been achieved precisely at $w=1$ using an additional constraint.

The invariant action \eqref{Action S3} can be seen as the non-relativistic analogue to the gravity coupled version of the relativistic CFT$_3$ given in eq.~\eqref{eq: CFT3}. The fact that, unlike in the relativistic case, we have another invariant  \eqref{Action S4}, on top of \eqref{Action S2} and \eqref{Action S3}, is specific to the fact that we are considering a complex scalar field. This is apparent from the classification in appendix \ref{ap: Kin term}.
This additional invariant can be obtained from the partially integrated Schr\"odinger action \eqref{Action S1}. We indicate it with a prime to distinguish it from the original Schr\"odinger action:
\begin{equation}\label{Action S1 bis}
S^{(5^\prime)}\equiv \int dtd^d{\bf x}\, e\, \left(i \Psi^\star D_0 \Psi - i \Psi D_0 \Psi^\star + \tfrac{1}{{\mass}} D_a\Psi^\star D_a \Psi \right)\,.
\end{equation}
This elucidates the relation between the action  \eqref{Action S4} and the Schr\"odinger action, namely,
the Lagrangian corresponding to the action \eqref{Action S4} is nothing else than  the square of the Schr\"odinger Lagrangian \eqref{Action S1 bis}. Note that the Lagrangian \eqref{Action S1 bis} is manifestly real. An important difference between the variation of the Lagrangians in equations \eqref{Action S1} and \eqref{Action S1 bis} is that the special conformal transformations fix $w=-\tfrac{d}{2}$ in the former case, see eq.~\eqref{eq: variation BoxSch}, whereas they do not fix the dilatation weight in the second case. Explicitly one finds $\delta  L_{(5^\prime)} = 2({w}-1)\Lambda_D  L_{(5^\prime)}$, with
\begin{equation}\label{eq: def Lag prim}
 L_{(5^\prime)}\equiv e^{-1}\mathcal L_{(5^\prime)}=i \Psi^\star D_0 \Psi - i \Psi D_0 \Psi^\star + \tfrac{1}{{\mass}} D_a\Psi^\star D_a \Psi\,.
\end{equation}
As a consequence, the Lagrangian  \eqref{Action S1 bis} can be squared, or taken to any higher power, to construct  invariant actions at higher orders with $w\neq-\tfrac{d}{2}$. An example of such a construction will follow below. In subsection \ref{gaugefixing}, we will explicitly see that after gauge fixing the coupled scalar field theories \eqref{Action S2}, \eqref{Action S3} and \eqref{Action S4} lead to three different Galilean invariants.

Following the classification of appendix \ref{ap: Kin term} we finally consider the SFT's that mix one time and two spatial derivatives. There are two such theories, see eqs. \eqref{Sch action 5} and \eqref{Sch action 6}. They are Schr\"odinger invariant for a complex scalar field of dilatation weight $w=-\tfrac{d-2}{2}$, as follows from the general expression \eqref{eq: nonrel weight}. They can be coupled to Schr\"odinger gravity in a straightforward way\begin{subequations}\label{mixL5prim}
\begin{align}
S^{(9)} &= \int dtd^d{\bf x}\, e\,\varphi^{-2} L_{(5^\prime)}\,D_a \varphi D_a \varphi \,,\label{Action S5} \\
S^{(10)} &= \int dtd^d{\bf x}\, e\,\varphi^{-1}  L_{(5^\prime)}\,\triangle \varphi \,.  \label{Action S6}
\end{align}
\end{subequations}
The invariance of actions \eqref{Action S5} and \eqref{Action S6} follows from the fact that  they are combinations of the Lagrangian \eqref{eq: def Lag prim} together with the building blocks \eqref{ActionPPP} that were used in the construction of the potential terms. We will see that although from the scalar field point of view these theories have a time derivative, after gauge fixing in subsection \ref{gaugefixing}, the invariants \eqref{Action S5} and \eqref{Action S6} will actually lead to Galilean potential terms containing only spatial derivatives.

\subsubsection{The curvature terms}\label{sec: Coupled Cuv}
As mentioned in section \ref{relcons}, in the relativistic case, there exist Poincar\'e invariants that do not arise from the coupling of scalar CFT's to conformal gravity. Here, in the same way, we expect that there exist locally Schr\"odinger invariants that are not related to any of the SFT's constructed in appendix \ref{Sch SFT}. Indeed, noticing that the Sch\"odinger spatial rotation curvature $R_{abcd}(J)$, defined in eq.~\eqref{Schcurvature}, only transforms under rotations and dilatations (with weight 2) and is invariant under the rest of the Schr\"odinger symmetries, it is clear that we can construct invariants out it. On the other hand, due to the fact that $R_{abcd}(J)=0$ in the flat space limit, it is also obvious that these invariants cannot arise from a scalar field theory in flat space.

In this section we build all possible locally Schr\"odinger invariant actions for the compensating scalar $\Psi$, with fixed dilatation weights $w= -\frac{d}{2}$ and $w= -\frac{d-2}{2}$, that are built out of curvatures of  Schr\"odinger gravity. As it turns out, only $\varphi$, the norm of $\Psi$, will actually appear in the invariants. In the construction of Schr\"odinger gravity most curvatures are set to zero as constraints in order to solve for the dependent gauge fields \eqref{dependentgaugefields} and \eqref{eq:vfz=2}. The only non-vanishing curvatures remaining are some components of the Schr\"odinger spatial rotation, Galilean boost and special conformal transformation curvatures, see \cite{Bergshoeff:2014uea} for the full details. However, it can be seen that all the invariants that can be built using the Galilean boost and special conformal transformation curvatures as well as $R_{0abc}(J)$ do necessarily break time-reversal symmetry. This is the reason why all the invariants we consider below are constructed only out of $R_{abcd}(J)$.

We start by enumerating the invariants that involve $R(J)\equiv R(J)_{ab}{}^{ab}$. At second order in spatial derivatives we can form the unique invariant
\begin{align}
S^{(11)} &= \int dtd^d{\bf x}\, e\, R(J) \, \varphi^2 \,,\qquad\qquad w= -\tfrac{d}{2}\,.\label{Action XI}
\end{align}
At higher order, it is  possible to combine $R(J)$ with itself, or together with $D_a\varphi D_a \varphi$, $\triangle \varphi$ and $L_{(5^\prime)}$ to form new invariants. We recall that $L_{(5^\prime)}$ has been defined in eq.~\eqref{eq: def Lag prim}. We summarize the invariants constructed in this way by giving the corresponding Lagrangians in the table below:
\vspace{0.25cm}
\begin{center}
\def\arraystretch{1.4}%
\begin{tabular}{ c|ccccc }
  $w=-\tfrac{d-2}{2}$ \,& $S^{(12)}$ & $S^{(13)}$ & $S^{(14)}$ & $S^{(15)}$ \\     \hline\hline
  $e^{-1}\mathcal{L}$ \,& \, $R(J)^2 \varphi^2$ \, & \, $D_a\varphi D_a\varphi R(J)$ \, & \, $\varphi \triangle \varphi R(J)$ \, & \, $L_{(5^\prime)} R(J) \,\varphi^2$ \\
\end{tabular}
\captionof{table}{This table indicates the four curvature  invariants $S^{(12)} - S^{(15)}$ that can be constructed using the contracted spatial rotation curvature $R(J)$.}
\end{center}
\vspace{0.25cm}

Of course, we can also construct additional invariants using other contractions of the Schr\"odinger rotation curvature tensor $R_{abcd}(J)$. This allows us to write three more invariants that we  summarize in the table below:
\vspace{0.25cm}
 \begin{center}
 \def\arraystretch{1.4}%
 \begin{tabular}{ c|cccc }
  $ w=-\tfrac{d-2}{2}$ \, & $S^{(16)}$ & $S^{(17)}$ & $S^{(18)}$ \\     \hline\hline
  $e^{-1}\mathcal{L}$ \, & \, $D_a\varphi D_b\varphi R^{ab}(J)$ \, & \, $\varphi^2R_{ab}(J)R^{ab}(J)$ \, & \, $\varphi^2R_{abcd}(J)R^{abcd}(J)$ \\
  \end{tabular}
  \captionof{table}{This table indicates the three additional curvature invariants $S^{(16)} - S^{(18)}$
  that can be constructed using the spatial rotation curvature $R^{ab}(J)$ or $R^{abcd}(J)$.}
  \end{center}
  \vspace{0.25cm}

Note that a Lagrangian of the form $\varphi D_aD_b\varphi R^{ab}(J)$ does not lead to a new independent invariant. It only differs from a combination of the invariant actions $S^{13}$, $S^{14}$ and $S^{16}$ by a boundary term due to the identity
\begin{align}
\varphi D_aD_b\varphi R^{a b}(J) &=  \frac{1}{2}\varphi \triangle \varphi R(J)  - D_a \varphi D_b\varphi R^{ab}(J) + \frac{1}{2}D_a\varphi D_a\varphi R(J) \nonumber\\
 & \quad + e^{-1}\partial_\mu\left[e \varphi e_{a}^{\mu}D_b\varphi\left( R^{a b}(J) - \frac{1}{2}\delta^{ab} R(J)\right)\right]\,.
\end{align}
These are all locally Schr\"odinger invariant actions built out of the Schr\"odinger curvatures at this order.

\subsection{Gauge fixing}\label{gaugefixing}
The scalar actions we have constructed so far are all locally invariant under Schr\"odinger symmetries. In this subsection we impose the gauge-fixing condition \eqref{gfc4} on the invariant actions constructed in section \ref{sec: Coupled SFT} to obtain the corresponding Galilean invariants.
To do so, we extensively make use of the formulas of appendix \ref{App A} in order to rewrite all Schr\"odinger dependent gauge fields in terms of purely Galilean quantities. At the end of this section we summarize in the tables \ref{Summary invariant result1} and \ref{Summary invariant result2} the  independent Galilean invariants that correspond to the actions $S^{(0\text{-}10)}$ and $S^{(11\text{-}18)}$, respectively.

We start by gauge fixing the invariant action $S^{(0)}$ which, upon fixing $\varphi=1$, trivially leads to a cosmological constant
\begin{equation}\label{Gal 1}
\mathrm{Gal}_0: \;\;e^{-1}{\mathcal{L}}_{\text{G} }= \Lambda_0 \,.
\end{equation}
Next we gauge-fix the invariant actions $S^{(1\text{-}4)}$ that are related to potential terms. These actions will contain  the same number of derivatives before and after gauge fixing. The action $S^{(1)}$, see eq.~\eqref{Action P1}, is invariant for a scalar field of weight $w=-\frac{d}{2}$. Using the definition of the covariant derivative \eqref{eq: def Davarphi}, we obtain after gauge fixing the following Galilean invariant:
\begin{equation}\label{Gal 1}
\mathrm{Gal}_1: \;\;e^{-1}{\mathcal{L}}_{\text{G} }=w^2 \,b\cdot b\,,
\end{equation}
where we recall that the dot refers to the contraction of the spatial indices. In the same way, we get for the higher-order potential terms,
\begin{align}
\mathrm{Gal}_2:  \;\; e^{-1}{\mathcal{L}}_{\text{G} }&= w^4 (b\cdot b)^2 \,,\label{Gal 2}\\
\mathrm{Gal}_3:  \;\;  e^{-1}{\mathcal{L}}_{\text{G} }&= -w^3(b\cdot b)(\cD\cdot b + w\,b\cdot b)\,,\label{Gal 3}\\
\mathrm{Gal}_4:  \;\;  e^{-1}{\mathcal{L}}_{\text{G} }&=w^2(\cD\cdot b + w\,b\cdot b)^2\,,\label{Gal 4}
\end{align}
with $w=-\frac{d-2}{2}$, and where the curly $\mathcal{D}$ is the Galilean covariant derivative, see eq.~\eqref{def. Galcovder1} for its definition. Note that although $\cD\cdot b$ is by itself a Galilean invariant, it does not lead to an independent invariant at second order due to the fact that $e(\cD\cdot b+2b\cdot b)$ is exactly a boundary term, see eq. \eqref{eq: Db bndy term}. This is consistent with what has been found from the local SFT point of view.

We continue with the prime example of a one-derivative kinetic term which is the  Schr\"odinger action \eqref{Saction} that has been coupled to Schr\"odinger gravity in eq.~\eqref{Action S1}. By inserting the explicit form of the dependent gauge fields in \eqref{Action S1} and applying the gauge fixing condition $\Psi=1$, we obtain the following Lagrangian
\begin{align}
\mathrm{Gal}_5:  \;\;  e^{-1}{\mathcal{L}}_{\text{G}}
&=\mass\Phi + \frac{1}{2\mass}w^2b\cdot b\,, \label{Gal 5}
\end{align}
with $w=-\frac{d}{2}$ and where $\Phi$ is defined in eq. \eqref{eq: def Phi}. After gauge-fixing to a frame with constant acceleration, the field $\Phi$ can be identified with the Newton potential. Note that the term $b_0$ has completely disappeared from the final result as expected from the invariance under special conformal transformations.

It turns out that by itself the Galilean invariant that is dual to the Schr\"odinger action \eqref{Saction} has inconsistent equations of motion.\footnote{This may be compared to the relativistic case where the cosmological constant by itself has an inconsistent equation of motion but it may be added in a consistent way to the Einstein-Hilbert action where it leads to a modification of the Einstein equations. See also footnote \ref{ftn}.} However, it can consistently be added to the Ho\v rava-Lifshitz action as we will discuss in section \ref{Relation to HLgravity}. Furthermore, it also shows up in the construction of the equations of motion for Newton-Cartan gravity  in section \ref{NC}.
As we saw above, $b\cdot b$ is an independent invariant which is generated by $S^{1}$, see eq.~\eqref{Gal 1}. Hence, $\Phi$ is the only independent Galilean invariant produced by the Schr\"odinger action.

We just showed that the Schr\"odinger action is unable to produce a kinetic term for the gravitational theory obtained after gauge fixing. This situation changes if we consider SFT's which are  second order in the time derivatives. In that case a kinetic term can be generated. The coupled complex scalar field theories $S^{(6\text{-}8)}$ given in eqs.~\eqref{2ndorderschrgr} are the only ones that are Schr\"odinger invariant with second order time derivatives, see appendix \ref{Sch SFT}. We will see that after gauge fixing the invariants $S^{(6\text{,}7)}$ will produce kinetic terms whereas the Galilean invariant produced by $S^{(8)}$ will not. Instead, the presence of this last term will be related to whether the field $\Phi$ can be integrated out from the final theory or not. We will come back to this point in section \ref{Relation to HLgravity}.

Upon gauge fixing $\Psi=1$ in the actions \eqref{2ndorderschrgr}, using $w=-\frac{d-2}{2}$ and after removing the boundary terms, we produce the following Galilean invariants,\begin{subequations}\label{galinvarinatsHL}
\begin{align}
\mathrm{Gal}_6:  \;\; e^{-1}\mathcal{L}_{\text G}
&=\frac{1}{d}K^{ab}K_{ab}-\frac{(w+1)}{2d}(K^a{}_a)^2+2w\Phi\left(\left(1+d^{-1}\right)\cD\cdot b +w b\cdot b\right) \nonumber\\[.1truecm]
& \qquad \qquad \qquad \qquad \qquad \qquad +\mass^2\left(\Phi-\frac{w}{2 \mass^2} \left(\mathcal D\cdot b+w b\cdot b\right)\right)^2\,,\label{Gal 6}\\[.1truecm]
\mathrm{Gal}_7:  \;\; e^{-1}\mathcal{L}_{\text{G}} &=  \frac{w^2}{d^2}(K^a{}_a)^2+ \mass^2\left(\Phi-\frac{w^2}{d\,\mass^2}\left(\cD\cdot b +(w+1)\,b\cdot b\right)\right)^2 \,,\label{Gal 7}\\[.1truecm]
\mathrm{Gal}_8:  \;\; e^{-1}\mathcal{L}_{\text{G}} &= \mass^2\Big(2 \Phi + \frac{w^2}{\mass^2}b\cdot b\Big)^2\,,\label{Gal 8}
\end{align}
\end{subequations}
where $K_{ab}$ has been defined in eq. \eqref{eq: hat Omegaab}. In section \ref{Relation to HLgravity} we will show that the terms $K^{ab}K_{ab}$ and $(K^a{}_a)^2$ correspond to the kinetic terms of $z=2$ Ho\v rava-Lifshitz Gravity.\footnote{$K_{ab}$ can be seen to correspond to a kinetic term due to the presence of the Galilean boost connection $\Omega_\mu^a$, see equations \eqref{Barg gauge fields2} and \eqref{def. Galcovder2} of appendix \ref{App A}.} As expected,  the $b_0$ contributions have  canceled out in the final results as a consequence of the  invariance under special conformal transformation. After computing the Galilean invariants $\mathrm{Gal}_{9}$ and $\mathrm{Gal}_{10}$, see below,  we will conclude that $K^{ab}K_{ab}$, $(K^a{}_a)^2$ and $\Phi^2$ are  the  three independent Galilean invariants that are generated by the actions $S^{(6\text{-}8)}$.

The invariant actions $S^{(9\text{-}10)}$, given by eqs.~\eqref{mixL5prim}, are first order in time derivative and second order in spatial derivatives. After gauge-fixing they only preserve their number of spatial derivatives. This is signaled by the presence of the vector field $M_\mu$ in the definition of $\Phi$. To be concrete, we obtain in this case, for $w=-\tfrac{d-2}{2}$,\begin{subequations}\label{mixGalprim}
\begin{align}
\mathrm{Gal}_9:  \;\; e^{-1}{\mathcal{L}}_{\text{G} }&= \mass w^2 \, (b\cdot b) ( 2\Phi + \frac{w^2}{\mass^2} \, b\cdot b ) \,,\label{Gal 9}\\
\mathrm{Gal}_{10}:  \;\; e^{-1}{\mathcal{L}}_{\text{G} }&= -\mass w(\cD\cdot b + w\,b\cdot b)( 2\Phi + \frac{w^2}{\mass^2} \, b\cdot b )\,,\label{Gal 10}
\end{align}
\end{subequations}
producing the new independent Galilean invariants $\Phi\, b\cdot b$ and $\Phi\,\cD\cdot b$.

For the convenience of the reader, we summarize in Table \ref{Summary invariant result1} below the independent Galilean invariants that we derived so far for each of the corresponding scalar SFT's after coupling them to Schr\"odinger gravity and performing a gauge fixing.

\vspace{0.2cm}
\begin{center}
\def\arraystretch{1.4}%
\begin{tabular}{c|ccccccccccc}
SFT \kern-0.3em & \kern-0.5em $(0)$ \kern-0.4em & \kern-0.5em $(1)$ \kern-0.4em & \kern-0.3em $(2)$ \kern-0.3em & \kern-0.3em $(3)$ \kern-0.2em & \kern-0.1em $(4)$ \kern-0.3em &\kern-0.3em $(5)$ \kern-0.3em & \kern-0.3em $(6)$ \kern-0.3em & \kern-0.3em $(7)$ \kern-0.3em & \kern-0.3em $(8)$ \kern-0.3em & \kern-0.2em $(9)$ \kern-0.3em & \kern-0.3em $(10)$ \kern-0.3em \\     \hline\hline
$\text{Gal. Inv.}$ \kern-0.2em & \kern-0.1em $\Lambda_0$ \kern-0.1em & \kern-0.1em $b^2$ \kern-0.1em & \kern-0.5em $b^4$ \kern-0.5em & \kern-0.2em $b^2 \mathcal{D}\cdot b$ \kern-0.2em & \kern-0.2em $(\mathcal{D}\cdot b)^2$ \kern-0.2em & \kern-0.5em $\Phi$ \kern-0.5em & \kern-0.2em $K^{ab}K_{ab}$\kern-0.2em &\kern-0.2em $(K^a{}_a)^2$ \kern-0.2em&\kern-0.3em $\Phi^2$ \kern-0.3em & \kern-0.2em $\Phi b^2$ \kern-0.2em & \kern-0.2em $\Phi\mathcal{D}\cdot b$\kern-0.2em \\
\end{tabular}
\captionof{table}{This table indicates the independent Galilean invariants that are produced by their corresponding SFT's. By $(0)$ in the first row we mean $S^{(0)}$, etc.}\label{Summary invariant result1}
\end{center}
\vspace{0.2cm}

Finally, we consider the Galilean invariants that are associated with the actions built out of the Schr\"odinger rotation curvature tensor $R(J)_{abcd}$ in subsubsection \ref{sec: Coupled Cuv}. For this purpose, we express $R(J)_{abcd}$ in terms of the Galilean rotation curvature tensor $\mathcal{R}_{abcd}(J)$, see its definition \eqref{Barg curvatures}. These two  curvatures are related by
\begin{align}
R(J)_{abcd} &= \mathcal{R}_{abcd}(J)
+ \cD_a b_{[d} \delta_{c] b} - \cD_b{b_{[d}}\delta_{c]a}
+ \cD_d{b_{[a}}\delta_{b] c} - \cD_c{b_{[a}}\delta_{b] d} \nonumber\\
&\quad + 2(b_{b}b_{[c}\delta_{d]a}-b_{a}b_{[c}\delta_{d] b}) + (b \cdot b)(\delta_{a c}\delta_{b d} - \delta_{a d}\delta_{b c})\,.\label{eq: relation Sch Gal Curvature}
\end{align}
Using expression \eqref{eq: relation Sch Gal Curvature}, we can rewrite the action $S^{(11)}$, see eq.~\eqref{Action XI}, after gauge-fixing as follows:
\begin{equation}
\mathrm{Gal}_{11}:  \;\; e^{-1}{\mathcal{L}}_{\text{G} }=\mathcal R +(d-1)\left(2\,\cD\cdot b-(d-2)\,b\cdot b\right)\,.
\end{equation}
The Galilean rotation curvature, $\cR=\mathcal{R}^{ab}{}_{ab}(J)$, is the only newly generated Galilean invariant. We recall that $\cD\cdot b \propto b\cdot b$ up to a boundary term.

Similarly, from expression \eqref{eq: relation Sch Gal Curvature} and the Galilean invariants that we already found, see table \ref{Summary invariant result1}, it can be seen that the actions $S^{(12\text{-}18)}$ produce Galilean invariants for which only the $\mathcal{R}_{abcd}$ contribution is leading to a new independent Galilean invariant. The remaining terms on the right hand side of eq.~\eqref{eq: relation Sch Gal Curvature} do not produce anything new. For simplicity, we only summarize the independent piece of the results in the table below.

\vspace{0.2cm}
\begin{center}
\def\arraystretch{1.4}%
\begin{tabular}{c|cccccccc}
  Curv. terms & \kern-0.1em $(11)$\kern-0.1em & \kern-0.4em $(12)$ \kern-0.4em & \kern-0.3em $(13)$ \kern-0.3em & \kern-0.2em $(14)$ \kern-0.2em & \kern-0.4em $(15)$ \kern-0.4em & $(16)$ & $(17)$ & $(18)$ \\     \hline\hline
  Gal. Inv. & \kern-0.1em $\mathcal{R}$ \kern-0.1em & \kern-0.4em $\mathcal{R}^2$ \kern-0.4em & \kern-0.3em $b^2\mathcal{R}$\kern-0.3em & \kern-0.2em $(\mathcal{D}\cdot b)\mathcal{R}$ \kern-0.2em &\kern-0.4em $\Phi \mathcal{R}$ \kern-0.4em & $b_ab_b\mathcal{R}^{ab}$ & $\mathcal{R}^{ab}\mathcal{R}_{ab}$ &  $\mathcal{R}^{abcd}\mathcal{R}_{abcd}$ \\
\end{tabular}
\captionof{table}{This table indicates the Galilean invariants that are produced by the curvature terms discussed in subsubsection 4.1.3. The (11) in the first row refers to the corresponding invariant action $S^{(11)}$, etc. }\label{Summary invariant result2}
\end{center}
\vspace{0.2cm}

This finishes our discussion of the non-relativistic conformal method applied to the Schr\"odinger invariant actions to obtain a number of Galilean invariants. These invariants appear naturally
in the $z=2$ Ho\v rava-Lifshitz gravity action which we discuss in the following section \ref{Relation to HLgravity}. 

\subsection{Identification with Ho\v rava-Lifshitz gravity}\label{Relation to HLgravity}
In this section we closely follow \cite{Hartong:2015zia} where it was first shown that, by making the NC geometry dynamical, HL gravity is reproduced from an action containing a collection of higher-derivative Galilean invariants. These are precisely the same invariants that we have derived in subsection \ref{gaugefixing} using the non-relativistic conformal method. For the convenience of the reader, and to make this work self-contained, we briefly repeat some of the arguments of \cite{Hartong:2015zia} using our own notation.

To start with, the most general action that we can construct out of the Galilean invariants obtained in section \ref{gaugefixing} is of the form
\bea\label{GG}
S=\frac{1}{\kappa^2}\int dt d^d{\textbf x}\, e \left(K_{ab}K^{ab}-\lambda \left(K^a{}_a\right)^2+ \cV\right)\,,
\eea
where $\lambda$ is a parameter and where  the potential $\cV$ contains any combination of the potential terms summarized in Table \ref{Summary invariant result1} and Table \ref{Summary invariant result2}, namely
\begin{align}
\cV  &= \Lambda_0+\lambda_1 \Phi + \lambda_2\Phi^2 + \lambda_3b^2 + \lambda_4b^4 + \lambda_5b^2 \mathcal{D}\cdot b + \lambda_6(\mathcal{D}\cdot b)^2 + \lambda_7\Phi b^2 + \lambda_8\Phi\mathcal{D}\cdot b  \nonumber\\
& \quad + \lambda_9\mathcal{R}  + \lambda_{10}\mathcal{R}^2  + \lambda_{11}b^2\mathcal{R} + \lambda_{12}(\mathcal{D}\cdot b)\mathcal{R} + \lambda_{13}\Phi \mathcal{R} \nonumber\\
& \quad + \lambda_{14}b_ab_b\mathcal{R}^{ab} + \lambda_{15}\mathcal{R}^{ab}\mathcal{R}_{ab} + \lambda_{16}\mathcal{R}^{abcd}\mathcal{R}_{abcd}\,,
\end{align}
for arbitrary coefficients $\lambda_1,\dots,\lambda_{16}$ and $\Lambda_0$. Note that if the coefficient  $\lambda_2$, in front of the $\Phi^2$ term, is non-vanishing then the field $\Phi$ can be integrated out. On the other hand, if $\lambda_2=0$ the equations of motion for $\Phi$ lead to a constraint equation.

We first identify our kinetic terms with the standard HL kinetic terms \cite{Horava:2009uw,Horava:2008ih}. In order to make the connection between our formalism and the one usually used in HL gravity we first rewrite the kinetic terms in terms of the spatial degenerate metric $h_{\mu\nu}\equiv \delta_{ab}e_\mu^ae_\nu^b$ and remove all the occurrences of the spatial vielbein $e_\mu^a$. Using the definition of $K_{ab}$ given by eq.~\eqref{eq: hat Omegaab}  we find that
\begin{align}
K_{ab} = e_a^\mu e_b^\nu\left(\frac{1}{2} \mathcal{L}_\tau(h_{\mu \nu}) + \frac{1}{2}\nabla_{\mu}( h_{\nu}{}^\rho M_\rho) + \frac{1}{2}\nabla_{\nu}( h_{\mu}{}^\rho M_\rho)  + 2M_{(\mu} b_{\nu)} \right)\,,\label{eq: hatOmega 2}
\end{align}
where
\begin{align}
\mathcal{L}_\tau(h_{\mu \nu}) &= \tau^\rho\partial_\rho h_{\mu\nu} + h_{\mu\rho}\partial_\nu \tau^\rho  + h_{\rho\nu}\partial_\mu \tau^\rho \,,\\
\nabla_\mu M_\nu &= \partial_{\mu} M_\nu - \Gamma^\rho_{\mu\nu} M_\rho\,,\\
\Gamma^\rho_{\mu\nu} &= \frac{1}{2}h^{\rho\sigma}\left(\partial_{\mu}h_{\nu\sigma}+\partial_{\nu}h_{\mu\sigma}+\partial_{\sigma}h_{\mu\nu}\right)\,,\label{eq: Gamma con}
\end{align}
with $h_{\mu}{}^\nu$ the spatial projector $h_{\mu}{}^\nu = h_{\mu\rho}h^{\nu\rho}$. Note that the expression \eqref{eq: Gamma con} is the only part of the Christoffel connection that is appearing due to the overall contraction with $e_a^\mu e_b^\nu$ in \eqref{eq: hatOmega 2}. Here, this $\Gamma^\rho_{\mu\nu}$ is merely used to obtain a rewriting of our kinetic terms in a purely metric formulation and should not be seen as a full connection. From the expression \eqref{eq: hatOmega 2}, it can then be checked that using a foliated ADM decomposition\footnote{We split the $\mu$ index into the coordinates $t$ and $x^i$.} with lapse and shift variables $N$ and $N_i$, respectively we have
\begin{equation}\label{eq: ADM}
\tau_\mu = \left(
\begin{array}{c}
N \\
0
\end{array}
\right)\,,\qquad
h_{\mu\nu} = \left(
\begin{array}{cc}
0&0\\
0&\gamma_{ij}
\end{array}
\right)\,, \qquad M_\mu = \left(
\begin{array}{c}
N M_0 \\
-N_i/N
\end{array}
\right)\,,
\end{equation}
we can rewrite our  kinetic terms in the form $K^{ab}K_{ab}=K^{ij}K_{ij}$ and $K^a{}_{a}=\gamma^{ij}K_{ij}$ with
\begin{equation}
K_{ij} = \frac{1}{2} N^{-1}\left(\partial_t\gamma_{ij}-\nabla_i N_j-\nabla_j N_i\right)\,,\label{eq: extrin curv}
\end{equation}
where $\nabla_i$ contains the standard Christoffel connection with respect to the invertible metric $\gamma_{ij}$. This is exactly the extrinsic curvature of HL gravity, and hence this shows the identification of the kinetic terms. For a more detailed dictionary between the Newton-Cartan and HL formalisms we refer the reader to \cite{Hartong:2015zia}.

The lapse field $N$ is the gauge field associated with time reparametrizations. In projectable HL gravity this gauge field  is restricted to be a projectable function on the spacetime foliation to preserve the time foliation, i.e.~$\partial_a N=0$. Using the expression \eqref{spatialb} for $b_a$ it follows that this corresponds to the  zero torsion case where $b_a=0$. This condition reduces the number of potential terms in the theory.

The HL action \eqref{GG} that we obtain differs from the one obtained in \cite{Hartong:2015zia} by the fact that it is not restricted to $d=2$. We obtain the result \eqref{GG} for arbitrary $d>0$.\footnote{As mentioned in appendix \ref{Sch SFT}, the case $d=2$ is included in our analysis although, by fixing the dilatation weight of the compensating scalar according to \eqref{eq: nonrel weight} for simplicity, we do not show this explicitly.} However, in the present setup, our analysis is restricted to $z=2$ instead. Moreover, because we consider SFTs only up to second order in time derivatives and fourth order in spatial derivatives, we reproduce the HL gravity potential considered e.g. in \cite{Blas:2010hb,Blas:2009qj,Sotiriou:2009gy} only up to the higher order derivative terms containing more than four spatial derivatives.



\section{Conclusions}\label{conclusion}

The relativistic conformal method has turned out to be very useful in many supergravity constructions.
In this paper we have developed a non-relativistic analogue of this formalism where the conformal algebra has been replaced by the smallest conformal extension
of the Bargmann algebra, i.e.~the $z=2$ Schr\"odinger algebra. The method guarantees that for each $z=2$ SFT one can construct a Galilean invariant. In this way one has a systematic way of constructing  all Galilean invariants of a given type except for the so-called curvature terms that do not correspond to a SFT.

In this paper we applied the non-relativistic conformal method to the SFT's with upto two time and four spatial derivatives. In this way we obtained a number of higher-derivative Galilean invariants that could be identified with $z=2$ Ho\v{r}ava-Lifshitz gravity thereby reproducing the results of \cite{Hartong:2015zia}. We expect that the non-relativistic conformal method will come to its full power once one wishes to study more complicated cases such as the supersymmetric extension of HL gravity. A first step in this direction has been taken in \cite{Bergshoeff:2015ija} following on the development of Newton-Cartan supergravity \cite{Andringa:2013mma,Bergshoeff:2015uaa}. Moreover, the classification of the higher order SFT's performed in appendix \ref{Sch SFT} is interesting in its own right and, in a different context, it could potentially be useful for other applications.

We also applied the non-relativistic conformal method to construct the equations of motion of curved space NC gravity with torsion, a result that, as far as we know, has not appeared before in  the literature. In this case, it was clearly an advantage to first extend the underlying $z=2$ SFT to the case with non-zero torsion instead of doing this straight-away in the NC equations of motion themselves. A peculiar feature of this construction is that we had to work with a SFT that was only defined in terms of equations of motion that, with the given number of fields, could not be integrated to an action. This reflects the property of NC gravity itself which is only formulated in terms of equations of motion without a clear underlying action principle. This is in contrast to the case of HL gravity considered in this paper which has an underlying action principle.

The formalism we developed in this work is naturally formulated in arbitrary dimensions. The generalization to values $z\ne 2$  of the dynamical exponent
is less obvious. A
necessary ingredient in achieving this is  to first construct Schr\"odinger gravity for $z\ne 2$. Fortunately, this has been already done in
\cite{Bergshoeff:2014uea}. It would be interesting to continue this program and construct the relevant SFT's. We expect that the analogy with the relativistic conformal programme will be less obvious due to the absence of the special conformal transformations for $z\ne 2$.

Finally, we note that the Schr\"odinger symmetries cannot be obtained as the non-relativistic limit of the conformal symmetries. Instead, one obtains the Galilean conformal symmetries that have also occurred in studies of non-relativistic holography \cite{Bagchi:2009my}. These symmetries are truly conformal in the sense that they do not allow a mass parameter. It would be interesting to see whether the non-relativistic conformal method can be extended to these
Galilean conformal symmetries as well.

\section*{Acknowledgements} We are grateful to Mehmet \"Ozkan for his many clarifications on the conformal method. We also thank Jan Rosseel, Charles Melby-Thompson and Thomas Zojer  for useful discussions on non-relativistic gravity. We thank Quim Gomis, Jelle Hartong and  Niels Obers 
for comments about an earlier version of this manuscript.
HA was supported by the Dutch stichting voor Fundamenteel Onderzoek der Materie (FOM). His work is supported in part by Iranian National Science Fundation (INSF). BR is supported by the Dutch stichting voor Fundamenteel Onderzoek der Materie (FOM). AM and PP were supported by Erasmus Mundus NAMASTE India-EU Grants. Some of the calculations have been performed using the software package Cadabra \cite{Peeters2007550}.

\appendix

\section{Schr\"odinger gravity}\label{App A}
In this appendix we collect a few formulae related to Schr\"odinger gravity which is obtained by gauging the $z=2$ Schr\"odinger algebra \cite{Bergshoeff:2014uea}. In the main text, we will find it useful to express the Schr\"odinger quantities in terms of the gauge fields and curvatures of the Bargmann algebra. For this purpose, it is convenient to relate in this appendix the Schr\"odinger gauge fields and curvatures  to the Bargmann gauge fields and curvatures  describing the Newton--Cartan geometry \cite{Andringa:2010it}.

The $z=2$ Schr\"{o}dinger algebra in $d+1$ dimensions reads
\begin{eqnarray}\label{Schrodingeralgebra}
&&[D,P_a] = -P_a\,,\hskip 1.3truecm [D,H] = -2H\,,\hskip 1.2truecm  [H,G_a] = P_a\,,\hskip 1.3truecm [P_a,G_b] = \delta_{ab}N\,,\nonumber\\[.2truecm]
&& [D,G_a] = G_a\,,\hskip 1.5truecm [D, K] = 2K\,,\hskip 1.5truecm [K, P_a] = -G_a\,,\hskip 1.0truecm [H,K] = D\,, \nonumber\\[.2truecm]
&&[J_{ab}, P_c] = 2\delta_{c[a}P_{b]}\,, \hskip 0.6truecm  [J_{ab}, G_c] = 2\delta_{c[a}G_{b]}\,, \hskip 0.6truecm  [J_{ab}, J_{cd}] = 4\delta_{[a[c}\,J_{b]d]}\,.
\end{eqnarray}
The corresponding gauge fields and gauge parameters of each generator are given in the table below.
We split the transformation of a gauge field $A_\mu$ into a general coordinate transformation, with parameter $\xi^\mu$, and the other `standard' gauge transformations according to
\bea
\delta A_\mu=\delta_\xi A_\mu +\partial_\mu \epsilon +[A_\mu,\epsilon]\,.
\eea
All gauge fields transform as covariant vectors under general coordinate transformations, so we usually only refer to the standard  gauge transformation piece.
\begin{center}\label{table5}
  \begin{tabular}{ l l l  l l  l l  }
   \hline
    $H$ & $P_a$ & $G_a$ & $J_{ab}$ & $D$ & $K$ & $N$\\     \hline
    $\xi^0$ & $\xi^a$ & $\Lambda^a$ & $\Lambda^{ab}$ & $\Lambda_D$ & $\Lambda_K$ & $\sigma$\\
    $\tau_\mu$ & $e_\mu{}^a$ & $\omega_\mu{}^a$ & $\omega_\mu{}^{ab}$ & $b_\mu$ & $f_\mu$ & $m_\mu$\\
    \hline
  \end{tabular}
  \captionof{table}{This table indicates the generators, parameters and gauge fields of the $z=2$ Schr\"odinger algebra.}
\end{center}

The time-like vielbein $\tau_\mu$, spatial vielbein $e_\mu{}^a$, central charge gauge field $m_\mu$ and the temporal projection of the dilatation gauge field $b_0=\tau^\mu b_\mu$
are independent gauge fields whose transformation rules are given by
\begin{subequations}
	\label{gaugetrafoszis1}
\begin{align}
\delta\tau_\mu & =  2\Lambda_D\tau_\mu\,, \\[.1truecm]
\delta e_\mu{}^a & =  \Lambda^a{}_b e_\mu{}^b+\Lambda^a\tau_\mu+\Lambda_De_\mu{}^a\,, \\[.1truecm]
\delta b_0 & =  \partial_0\Lambda_D+\Lambda_K-\Lambda^a b_a - 2\Lambda_D  b_0\,,  \\[.1truecm]
\delta m_\mu & =  \partial_\mu\sigma+\Lambda^a e_{\mu a}\,.
\end{align}
\end{subequations}
We consider the case of twistless torsion, meaning that $\tau_\mu$ additionally satisfies
\begin{equation}
e^\mu_a e^\nu_b \left(\partial_\mu\tau_\nu - \partial_\nu\tau_\mu \right) = 0 \,.\label{eq: twistless}
\end{equation}
The remaining Schr\"odinger gauge fields are dependent and are solved for in terms of the independent ones by imposing curvature constraints whose formulae we do not give here.
The dependent gauge fields $\omega_\mu{}^a{}^b$, $\omega_\mu{}^a$, $b_a$ and $f_a$ are expressed in terms of $\tau_\mu$, $e_\mu^a$, $m_\mu$ and $b_0$ as follows:\begin{subequations}\label{dependentgaugefields}
\begin{align}
 \omega_\mu{}^a{}^b&=\Omega_\mu{}^a{}^b+2e_\mu{}^{[a} b^{ b]}\,,\label{Omegas1}\\[.2truecm]
 \omega_\mu{}^a&=\Omega_\mu{}^a+e_\mu{}^a b_0\,,\label{Omegas2}\\[.2truecm]
 b_a&=e_a{}^\mu\tau^\nu\pa_{[\mu}\tau_{\nu]}\label{spatialb}\,,\\[.2truecm]
 f_a &=2e_a{}^\mu\tau^\nu\partial_{[\mu} b_{\nu]}\,,\label{spatialf}
\end{align}
\end{subequations}
where
 \begin{align}
  \Omega_\mu{}^a{}^b&=-2e^\nu{}^{[a} \pa_{[\mu} e_{\nu]}{}^{b]}+e_\mu{}_ce^\nu{}^ae^\rho{}^b \pa_{[\nu} e_{\rho]}{}^c-\tau_\mu e^\nu{}^ae^\rho{}^b\pa_{[\nu} m_{\rho]}\,,\label{Barg gauge fields1}\\[.2truecm]
  \Omega_\mu{}^a&=\tau^\nu \pa_{[\mu} e_{\nu]}{}^a+\tau^\nu e^\rho{}^a e_\mu{}_b \pa_{[\rho} e_{\nu]}{}^{b}+e^\nu{}^a\pa_{[\mu} m_{\nu]}+\tau_\mu \tau^\rho e^\nu{}^a\pa_{[\rho} m_{\nu]}\,,\label{Barg gauge fields2}
 \end{align}
are the rotation and boost gauge fields of the Bargmann algebra. The remaining dependent gauge field $f_0$ cannot be solved for using a fully gauge invariant constraint written purely in terms of Schr\"odinger gauge fields, see the discussion in \cite{Bergshoeff:2014uea}. However, using $M_\mu$, a vector transforming under boosts as in eq.~\eqref{acc}, this problem can be circumvented and a gauge invariant constraint provides the solution:
\begin{align}
f_0  &=\frac{2}{d}\tau^\mu e^\nu{}_a\left(D_{[\mu}\omega_{\nu]}{}^a+b_{[\mu}\omega_{\nu]}{}^a\right)+\frac{2}{d}\left(R_{0 a}{}^{ac}(J)+\frac{1}{2}M^aR_{ab}{}^{bc}(J)\right)M_c\,,\label{eq:vfz=2}
\end{align}
where $D_\mu$ is the covariant derivative with respect to the Schr\"odinger spatial rotations and $R_{\mu\nu}{}^{ab}(J)$ is the Schr\"odinger spatial curvature defined in \eqref{Schcurvature}. Whenever in the text one of these dependent gauge fields occur, it is understood that they are given by the expressions above.

The transformation rules of the dependent gauge fields are given by
\begin{subequations}
	\label{gaugetrafoszis2}
\begin{align}
\delta\omega_\mu{}^{ab} & =  D_\mu\Lambda^{ab}\,, \\[.1truecm]
\delta\omega_\mu{}^a & =  D_\mu\Lambda^a+\Lambda^{a}{}_b\omega_{\mu}{}^b+\Lambda^ab_\mu-\Lambda_D\omega_\mu{}^a+\Lambda_K e_\mu{}^a\,,  \\[.1truecm]
\delta b_a & =  \partial_a\Lambda_D - \Lambda_D b_a + \Lambda_a{}^b b_b\,,  \\[.1truecm]
\delta f_\mu  &=  \partial_\mu\Lambda_K+2\Lambda_K b_\mu-2\Lambda_Df_\mu\,.
\end{align}
\end{subequations}
The Schr\"odinger curvatures that we use explicitly in this work are the curvatures $R(J)$ associated to spatial rotations and $R(G)$ corresponding to boosts. These curvatures are given by\begin{subequations}
	\label{Schcurvature}
\begin{align}	\label{SchcurvatureRJ}
R_{\mu\nu}{}^{ab}(J) & =  2\partial_{[\mu}\omega_{\nu]}{}^{ab}-2\omega_{[\mu}{}^{c[a}\omega_{\nu]}{}^{b]}{}_c\,, \\[.1truecm]
R_{\mu\nu}{}^a(G) & =  2\partial_{[\mu}\omega_{\nu]}{}^a+2\omega_{[\mu}{}^b\omega_{\nu]}{}^a{}_b-2\omega_{[\mu}{}^ab_{\nu]}-2f_{[\mu}e_{\nu]}{}^a\,.
\end{align}
\end{subequations}
We define the curvatures for the Bargmann rotation connection $\Omega_\mu{}^{ab}$ in \eqref{Barg gauge fields1} and for the Bargmann  boost gauge field $\Omega_\mu{}^a$ in \eqref{Barg gauge fields2} as follows:\begin{subequations}
	 \label{Barg curvatures}
\begin{align}
\mathcal{R}_{\mu\nu}{}^{ab}(J) & =   2\partial_{[\mu}\Omega_{\nu]}{}^{ab}-2\Omega_{[\mu}{}^{c[a}\Omega_{\nu]}{}^{b]}{}_c\,,  \label{Barg curvaturesJ}\\[.1truecm]
\mathcal{R}_{\mu\nu}{}^a(G) & =  2\partial_{[\mu}\Omega_{\nu]}{}^a+2\Omega_{[\mu}{}^b\Omega_{\nu]}{}^a{}_b\,. \label{Barg curvaturesG}
\end{align}
\end{subequations}
Formally, one can obtain the Bargmann quantities \eqref{Barg gauge fields1}, \eqref{Barg gauge fields2} and  \eqref{Barg curvatures} from the corresponding Schr\"odinger expressions  \eqref{Omegas1}, \eqref{Omegas2} and  \eqref{Schcurvature} by setting $b_\mu=f_\mu=0$.\,\footnote{We notice that in describing the Newton--Cartan geometry the vielbeine $(\tau_\mu )^{\text{G}}$ and $(e^a_\mu )^{\text{G}}$ are in general not the same as the Schr\"odinger
ones. They are only the same after gauge fixing, see eq.~\eqref{dc}.} We will also use the Galilean covariant derivatives
\begin{subequations}
\label{def. Galcovder012}
\begin{align}
\mathcal{D}_{0} b_a &= \tau^\mu\left(\partial_{\mu}b_a - \Omega_{\mu a}{}^{c}b_c\right)\,,\label{def. Galcovder0}\\[.1truecm]
\mathcal{D}_a b_b &= e_a^\mu\left(\partial_{\mu}b_b - \Omega_{\mu b}{}^{c}b_c\right)\,,\label{def. Galcovder1}\\[.1truecm]
\mathcal{D}_a M_b &= e_a^\mu\left(\partial_{\mu}M_b - \Omega_{\mu b}{}^{c}M_c - \Omega_{\mu b}\right) \,.\label{def. Galcovder2}
\end{align}
\end{subequations}
Note that $\mathcal{D}_a b_b$ is a Galilean boost invariant by itself. This is unlike $\mathcal{D}_{0} b_a$ and $\mathcal{D}_a M_b$ which can however be used to construct a Galilean boost invariant in conjunction with additional terms. We find the Galilean boost invariant combinations:
\begin{subequations}
\label{def. Galinv}
\begin{align}
K_a &= \mathcal{D}_{0} b_a + M^b\mathcal{D}_b b_a + b_ab_bM^b - M_a b\cdot b\,,\label{def. Galinv1}\\[.1truecm]
K_{ab} &= \mathcal{D}_a M_b +M_ab_b+M_bb_a\,.\label{def. Galinv2}
\end{align}
\end{subequations}
It follows from the definitions \eqref{eq: twistless} and \eqref{Barg gauge fields1} that $\mathcal{D}_{[a} b_{b]} =0$. Furthermore, from the same equations \eqref{eq: twistless} and \eqref{Barg gauge fields1} it can be seen that, as a consequence of having twistless torsion, the term $\mathcal{D} \cdot b\equiv \delta^{ab}\mathcal{D}_a b_b$ is a total derivative only up to an additional torsion contribution, namely:
\begin{equation}
\mathcal{D} \cdot b + 2\,b\cdot b = e^{-1}\partial_\mu\left(ee^\mu_ab^a\right)\,.\label{eq: Db bndy term}
\end{equation}
Using the definitions \eqref{Barg gauge fields1} and \eqref{Barg gauge fields2} one may also verify that the anti-symmetric part of $\mathcal{D}_a M_b$ is vanishing
\begin{align}\label{eq: antisym part DM}
\mathcal{D}_{[a} M_{b]} = e_a^\mu e_b^\nu \left( \partial_{[\mu}M_{\nu]} - \partial_{[\mu}m_{\nu]}\right) = 0\,.
\end{align}
To show the last step,  one  uses the relation between $M_\mu$ and $m_\mu$ given in \eqref{dc}.


\section{Scalar Schr\"odinger field theories}\label{Sch SFT}
In this appendix we classify all possible independent complex scalar field theories invariant under rigid $z=2$ Schr\"odinger transformations up to second order in time-derivatives and fourth order in spatial-derivatives.  Explicitly, the complex scalar field $\Psi$ transforms according to eq.~\eqref{deltapsi} with the parameters given by eqs.~\eqref{solution1} as follows:
\begin{align}
\delta\Psi&=\big[\xi^0\partial_0+\xi^a\partial_a+w\Lambda_D+i\mass\sigma\big]\Psi \nonumber\\
&=\big(a^0-2\lambda_Dt+\lambda_Kt^2\big)\partial_0\Psi +\big(a^c-\lambda^{cb}x_b-\lambda^ct-\lambda_Dx^c+\lambda_Ktx^c\big)\partial_c\Psi \nonumber\\
&\quad +w\big(\lambda_D-\lambda_Kt\big)\Psi+i\mass \big(\sigma_0-\lambda^ax_a+\frac{1}{2}\lambda_Kx^2\big)\Psi\,.\label{eq: ap transfo Psi}
\end{align}
Here all parameters are constants, $w$ is the dilatation weight and $\mass$ is the  weight under central charge transformations.

The independent scalar SFT's we will obtain are only defined up to boundary terms. However, because ultimately we are interested in coupling these scalar field theories to construct local invariants, we will fix this ambiguity by performing our classification directly at the level of the Lagrangian. We recall that, strictly speaking, the Lagrangians are never invariant under the transformation \eqref{eq: ap transfo Psi}. Hence, we will classify the Lagrangians by imposing that they transform according to the following total derivative:
\begin{equation}\label{eq: ap var L condition}
\delta \mathcal{L} = \partial_0 \left(\xi^0 \mathcal{L}\right) + \partial_a \left(\xi^a \mathcal{L}\right)\,.
\end{equation}
With slight abuse of terminology we will refer to a Lagrangian satisfying \eqref{eq: ap var L condition} as an invariant Lagrangian. This condition guaranties that the action will be invariant under the full Sch\"odinger transformations \eqref{eq: ap transfo Psi} and that the Lagrangian admits a coupling to Schr\"odinger gravity, see also the discussion just above and below equation \eqref{eq: Condition Inv Lag}. In general, the boundary terms generated by performing partial integrations are not invariant by themselves. Therefore, two Lagrangians that are related by such a boundary term will not both be invariant upon throwing away the boundary term and this fixes one preferred Lagrangian over the other. As we will see next, in one specific case the boundary term is guaranteed to be an invariant. In this case, the analysis can be performed directly at the level of the action.

The way we organize the classification goes as follow. We start in subsection \ref{ap: Pot term} with the classification of the potential terms, i.e.~terms without time-derivatives, and in subsection \ref{ap: Kin term} we consider the more involved case of Schr\"odinger invariant actions with time-derivatives. Depending on what is most convenient we will use a formulation where the derivatives act on the complex scalar $\Psi$ and its conjugate $\Psi^\star$ or on the two real scalars $\varphi$ and $\chi$. We recall that
\begin{equation}
\Psi = \varphi e^{i\chi}\,.
\end{equation}
We will first fix the most general terms that can be written down by requiring invariance under dilatation and central charge symmetry. In a second step we will add the required compensating terms to form invariants under Galilean boost and special conformal symmetries.

In principle the analysis can be done for arbitrary spatial dimensions $d$ and arbitrary dilatation weight $w$. However, it can be seen that the SFT's we obtain are significantly simplified by fixing the dilatation weight to
\begin{equation}\label{eq: app weight}
w = - \frac{d+2-2n_t-n_s}{2}\,,
\end{equation}
where $n_t$ is the number of time derivatives and $n_s$ the number of spatial derivatives. In order to improve the presentation of our results we will therefore fix the weight according to \eqref{eq: app weight} from the start. Note that by construction we are assuming $w\neq0$. Strictly speaking, the condition \eqref{eq: app weight} is therefore a condition on the spatial dimension. However, nothing depends crucially on the choice \eqref{eq: app weight} and it is straightforward to include the case $d=2n_t+n_s-2$ by fixing $w$ differently.

\subsection{Potential terms}\label{ap: Pot term}
We start our classification by the potential terms. In this case we write the complex scalar $\Psi$ in terms of its norm $\varphi$ and angle $\chi$. In a first step, the construction of invariant potential terms is simplified by letting the spatial derivatives act only on the norm $\varphi$. This case is very special: provided all indices are contracted we are guaranteed to produce an invariant. For that reason also the possible boundary terms created by partial integrations are invariants by themselves. This means that in this case it is sufficient to perform the classification at the level of the action. This first part of our analysis is therefore equivalent to the classification of all inequivalent ways to contract the spatial indices of the derivatives acting on $\varphi$ up to boundary terms. In a second step we look at possible invariants where a spatial derivative is also allowed to act on $\chi$. We will find that there are no such term with two spatial derivatives and only one at the next order with four derivatives.

At zeroth order, with $n_t=n_s=0$, we fix the dilatation weight to $-\tfrac{d+2}{2}$ according to \eqref{eq: app weight}. It follows that the only Schr\"odinger invariant action takes the form
\begin{equation}\label{eq: flat P0}
\mathrm{SFT}_{0}:\;\; S^{(0)} = \int dtd^dx \, \Lambda_0\varphi^2\,,
\end{equation}
where $\Lambda_0$ is an arbitrary function.

At second order ($n_t=0$, $n_s=2$), with spatial derivatives acting only on $\varphi$, the dilatation symmetry implies that we can only write down two terms in the Lagrangian: $\partial^a\varphi\partial_a\varphi$ and $\varphi\partial^a\partial_a\varphi$. All other symmetries are then automatically satisfied. Moreover, these two terms are related by an invariant boundary term, hence there is no difference between using one or the other.\footnote{In section \ref{HLgravity} we confirm that both terms do indeed lead to the same action after coupling to Schr\"odinger gravity.} Therefore, we find that the unique Schr\"odinger field theory at this order is
\begin{equation}\label{eq: flat P1}
\mathrm{SFT}_{1}:\;\; S^{(1)} = \int dtd^dx \, \partial^a\varphi\partial_a\varphi\,.
\end{equation}
We recall that the weight of the scalar field $\varphi$ is fixed by \eqref{eq: app weight}.

Performing the same analysis at fourth order in the spatial derivatives  ($n_t=0$, $n_s=4$), we find three independent invariant SFT's given by\begin{subequations}
		\label{eq: flat Ps}
\begin{align}
\mathrm{SFT}_{2}:\;\; S^{(2)} &= \int dtd^dx\, \varphi^{-2}(\partial^a \varphi \partial_a  \varphi)^2\,,\label{eq: flat P2}\\
\mathrm{SFT}_{3}:\;\; S^{(3)} &= \int dtd^dx\, \varphi^{-1}(\partial^a \varphi \partial_a  \varphi) (\partial^b\partial_b \varphi)\,,\label{eq: flat P3}\\
\mathrm{SFT}_{4}:\;\; S^{(4)} &= \int dtd^dx\, (\partial^a\partial_a \varphi)^2\,.\label{eq: flat P4}
\end{align}
\end{subequations}
These correspond to the only three possible actions that can be built with four spatial derivatives acting on $\varphi$ and that cannot be related by partial integrations.

Let us now analyze whether there are also invariant potential terms involving the real scalar field $\chi$. First of all, note that if we write a term where $\chi$ appears without any derivatives the Lagrangian cannot be made invariant under central charge symmetry. Furthermore, the scalar field $\chi$ is odd under time reversal. Because we consider only spacial derivatives we need $\chi$ to appear an even number of times in order to produce time reversal invariant field theories.

With two spatial derivatives the only term that we can write down with correct scaling behavior is: $\varphi^2\partial^a\chi\partial_a \chi$.
However, this term by itself is not an invariant and there is nothing else that can be written down that could be added to make it invariant. We conclude that \eqref{eq: flat P1} is the only invariant at second order in spatial derivatives.\footnote{This would also be the case even without imposing time reversal invariance.}

At fourth order the situation is more interesting. Imposing central charge, dilatation and time reversal symmetry, we are in principle allowed to consider the following list of terms:
\begin{align}
& \varphi\partial^a\partial_a\varphi \partial_b\chi\partial^b\chi\,,\; \varphi\partial_a\partial_b\varphi  \partial^a\chi\partial^b\chi\,,\; \partial^a\varphi\partial_a\varphi \partial_b\chi\partial^b\chi\,,\; \partial_a\varphi\partial_b\varphi\partial^a\chi\partial^b\chi \,,\nonumber\\
&\varphi\partial_a\varphi\partial^a \chi \partial^b\partial_b\chi\,,\; \varphi\partial_a\varphi\partial^a \partial^b \chi \partial_b\chi \,,\nonumber\\
&\varphi^{2}\partial_a\chi\partial^a\chi\partial_b\chi\partial^b\chi\,,\; \varphi^{2}\partial^a\partial_a\chi\partial_b\partial^b\chi\,,\; \varphi^{2}\partial^a\partial^b\chi\partial_a\partial_b\chi\,,\; \varphi^{2}\partial^a\partial_a\partial^b\chi\partial_b\chi\,,
\label{eq: terms with chi ns4}
\end{align}
respectively with two, three and four derivatives acting on $\chi$. Taking the most general linear combination of the terms in \eqref{eq: terms with chi ns4} we find a unique Schr\"odinger invariant given by
\begin{equation}
\mathrm{SFT}_{6'}:\;\; S^{(6')} = \int dtd^dx\,\varphi^{2}\left( (\partial^a\partial_a\chi)^2 - d  \partial_a\partial_b\chi \partial^a\partial^b\chi\right)\,.\label{eq: flat P5}
\end{equation}
Interestingly, although this is a potential term, after coupling to Schr\"odinger gravity and gauge fixing this invariant does generate a linear combination of the HL kinetic terms. However, we do not consider it in the main text where we prefer to work with another set of independent invariants generated by the complex field $\Psi$ (see section \ref{ap: Kin term}). This is the reason why we added a prime on the label. This will become clearer in the next section when we give the explicit relation between the invariant \eqref{eq: flat P5} and the ones we will use effectively. This concludes our discussion of the potential terms.

\subsection{Kinetic terms}\label{ap: Kin term}

We now turn to the construction of scalar SFT's that contain time derivatives. In this case we will mostly work with the derivatives acting on the fields $\Psi$ and $\Psi^\star$ in order to make contact with the Schr\"odinger action in its most well-known form. 

We start by looking at possible invariant theories containing a single time derivative ($n_t=1$, $n_s=0$) where, we can without loss of generality fix the dilatation weight according to eq. \eqref{eq: app weight} to $w=-\tfrac{d}{2}$. Imposing invariance under central charge and dilatation symmetry there are only two terms that we can write down with a single time derivative: $\Psi^\star\partial_0\Psi$ and its complex conjugate $\Psi\partial_0\Psi^\star$. Let us consider $\Psi^\star\partial_0\Psi$, by computing the variation \eqref{eq: ap transfo Psi} on this term we obtain
\begin{equation}\label{eq: var KinS5}
\delta \left(\Psi^\star\partial_0\Psi \right) \cong - \, \Psi^\star \left(\lambda^a\partial_a+\lambda_K (w-x^{a}\partial_a ) \right)\Psi\,,
\end{equation}
where by $\cong$ we mean equality up to the total derivative of equation \eqref{eq: ap var L condition}. The terms on the right hand side of \eqref{eq: var KinS5} can be exactly compensated for by the addition on the left hand side of extra terms containing only spatial derivatives. The resulting invariant combination is the Lagrangian describing the Schr\"odinger action,
\begin{equation}\label{Sch action}
\mathrm{SFT}_{5}:\;\; S^{(5)} = \int dtd^dx\, \Psi^\star\Big(i \partial_0  - \tfrac{1}{2{\text \mass }}\partial^a\partial_a\Big) \Psi \,.
\end{equation}
This is the action that we will couple to Schr\"odinger gravity in section \ref{sec: Coupled Kin}.

Although the Lagrangian of eq. \eqref{Sch action} has an imaginary part, the Schr\"odinger action itself is real. Performing a similar analysis starting with $\Psi\partial_0\Psi^\star$ we would just find the complex conjugate of the Lagrangian of equation \eqref{Sch action}, leading to nothing new. Therefore, we conclude that, up to invariant potential terms, the Schr\"odinger action \eqref{Sch action} is the unique invariant at first order in time derivative.

We next consider SFT's at second order in time derivatives and following \eqref{eq: app weight} we fix the dilatation weight to $w=-\tfrac{d-2}{2}$. Here the possibilities increase. Imposing dilatation and central charge invariance leads to the following five possible kinetic terms
\begin{equation}
\Psi^\star\partial_0^2 \Psi\,,\;\Psi\partial_0^2 \Psi^\star\,, \;\partial_0 \Psi \partial_0 \Psi^\star\,,\; \Psi^\star \Psi^{-1}(\partial_0 \Psi)^2\,,\; \Psi \Psi^\star{}^{-1}(\partial_0 \Psi^\star)^2\,.
\end{equation}
We now perform a similar analysis as in the first order case. Up to complex conjugation there are three cases to consider.
\bigskip

\noindent (1)\  \ For the term $\Psi^\star\partial_0^2 \Psi$, we find that its variation
\begin{equation}
\delta\left(\Psi^\star\partial_0^2 \Psi \right) \cong -2\Psi^\star\left(\lambda^a\partial_a\partial_0 + \lambda_K \left((w-1)\partial_0 - x^a\partial_a\partial_0\right) \right)\Psi
\end{equation}
can be exactly compensated for by the addition of extra terms. The corresponding invariant Lagrangian is leading to the action
\begin{align}
\mathrm{SFT}_{6}:\;\; S^{(6)}
&=\int dtd^dx\, \Psi^\star \left(i\partial_0-\frac{1}{2\mass}\partial^a\partial_a \right)^2\Psi\label{Sch action 2}\,,
\end{align}
which we recognize as the square of \eqref{Sch action}. For the same reasons as in the first order case, the complex conjugate $\Psi\partial_0^2 \Psi^\star$ cannot give anything new.

\bigskip

\noindent (2)\  \ The variation of the next possible term at this order, namely $\partial_0 \Psi \partial_0 \Psi^\star$, is
\begin{eqnarray}
\delta\left(\partial_0 \Psi \partial_0 \Psi^\star\right)   \cong  \left(\lambda_K x^a -\lambda^a \right)(\partial_a\Psi^\star  \partial_0\Psi +  \partial_a\Psi  \partial_0 \Psi^\star)  - w \lambda_K  (\Psi^\star\partial_0\Psi  +\Psi \partial_0\Psi^\star ) \,.
\end{eqnarray}
Again, the variation above can be compensated for leading to a unique invariant Lagrangian:
\begin{align}
\mathrm{SFT}_{7}:\;\; S^{(7)}  &=\int dtd^dx\, \left| \left(i\partial_0 -\frac{1}{2\mass}\partial^a\partial_a \right)\Psi + \frac{1}{\mass d}\left(\partial^a\partial_a \Psi - \frac{1}{\Psi} (\partial_a \Psi)^2\right) \right|^2\,,\label{Sch action 3}
\end{align}
where $\left| \cdot \right|$ is the norm.

\bigskip

\noindent (3) \  \ Proceeding in a similar way with the kinetic term $\Psi^\star \Psi^{-1}(\partial_0 \Psi)^2$ it turns out that its variation:
\begin{equation}
\delta\left(\Psi^\star \Psi^{-1}(\partial_0 \Psi)^2\right)  \cong -2\Psi^\star\partial_0\Psi\left(w\lambda_K + \Psi^{-1} \left(\lambda^a-x^a\lambda_K\right) \partial_a\Psi \right)\,,
\end{equation}
cannot directly be compensated for. In this case, it is necessary to combine this term with its complex conjugate. In order to simplify the final interpretation of this action we will find it convenient to actually consider the variation of the following combination of terms
\begin{equation}\label{eq: variation combin}
 - \Psi^\star \frac{(\partial_0 \Psi)^2}{\Psi} - \Psi\frac{(\partial_0 \Psi^\star)^2}{\Psi^\star} + 2\partial_0 \Psi \partial_0 \Psi^\star
=-\varphi^{-2} \left(\Psi^\star\partial_0\Psi - \Psi\partial_0\Psi^\star \right)^2,
\end{equation}
where an additional $\partial_0 \Psi \partial_0 \Psi^\star$ term has been added to complete the square. After compensating the variation of the combination in \eqref{eq: variation combin} we find the following invariant action
\begin{align}
\mathrm{SFT}_{8}:\;\; S^{(8)}&= \int dtdx^a  \frac{1}{\Psi^\star\Psi}\left( i\Psi^\star\partial_0 \Psi - i\Psi \partial_0 \Psi^\star +  \frac{1}{\mass }\partial_a \Psi \partial_a \Psi^\star \right)^2 \,.\label{Sch action 4}
\end{align}
We can recognize this action as the square of the partially integrated Schr\"odinger action, see also the discussion above \eqref{Action S1 bis} in the main text. This exhausts all the possibilities and concludes our discussion of the kinetic terms with two time derivatives.

At last, we look at possible invariant with one time and two spatial derivatives in terms of $\varphi$ and $\chi$. Imposing central charge, dilatation and time-reversal symmetry the only terms that can be written down are
\begin{eqnarray}
\varphi^2\partial_0\partial^2\chi \,,\; \varphi^2\partial_0\chi\partial^a\chi\partial_a\chi \,, \; \varphi\partial_0\varphi\partial^2\chi \,,\; \varphi\partial^a\varphi\partial_0\partial_a\chi \,, \nonumber\\
\varphi\partial_0\partial^a\varphi\partial_a\chi\,,\; \varphi\partial^2\varphi\partial_0\chi \,,\; \partial_0\varphi\partial^a\varphi\partial_a\chi \,,\; \partial^a\varphi\partial_a\varphi\partial_0\chi \,, \label{eq: inv term 12}
\end{eqnarray}
where we recall that the dilatation weight is fixed to $w=-\tfrac{d-2}{2}$. From the terms in \eqref{eq: inv term 12} it is possible to build three invariant Lagrangians:
\begin{subequations}\label{eq: mix Inv123}
\begin{eqnarray}
I_1&=&\partial^a\varphi\partial_a\varphi\partial_0\chi-\frac{1}{2\mass }(\partial^a\varphi\partial_a\varphi\partial^b\chi\partial_b\chi) \,,\label{eq: mix Inv1}\\
I_2&=&\varphi\partial_a\partial^a\varphi\partial_0\chi -\frac{1}{2\mass }(\varphi\partial_a\partial^a\varphi\partial^b\chi\partial_b\chi) \,,\label{eq: mix Inv2}\\
I_3&=&\varphi\partial^a\varphi\partial_0\partial_a\chi-\frac{1}{\mass }(\varphi\partial^a\varphi\partial^b\chi\partial_a\partial_b\chi)\,.\label{eq: mix Inv3}
\end{eqnarray}
\end{subequations}
However, it can be seen that these Lagrangians lead to only two linearly independent actions as $I_3$ is nothing but a combination of $I_1$ and $I_2$ up to a partial integration using $\partial_a$. The invariant Lagrangians \eqref{eq: mix Inv1} and \eqref{eq: mix Inv2} can be rewritten as the following invariant actions\begin{subequations}
	\label{Sch action 5,6}
\begin{align}
\mathrm{SFT}_{9}&:\;\; & S^{(9)}&=\int dtd^dx\, \frac{1}{\varphi^2}\partial^a\varphi\partial_a\varphi\left(i\Psi^\star\partial_0\Psi-i\Psi\partial_0\Psi^\star+\frac{1}{\mass }\partial^a\Psi^\star\partial_a\Psi\right)\,,\label{Sch action 5}\\
\mathrm{SFT}_{10}&:\;\; & S^{(10)}&=\int dtd^dx\, \frac{1}{\varphi}\partial^2\varphi\left(i\Psi^\star\partial_0\Psi-i\Psi\partial_0\Psi^\star+\frac{1}{\mass }\partial^a\Psi^\star\partial_a\Psi\right) \,,\label{Sch action 6}
\end{align}
\end{subequations}
where for convenience we have ignored an overall factor of $-\tfrac{1}{2}$ and added the invariant potential terms $\tfrac{1}{2}S^{(2)}$ and $\tfrac{1}{2}S^{(3)}$ to \eqref{eq: mix Inv1} and \eqref{eq: mix Inv2} respectively. The actions \eqref{Sch action 5} and \eqref{Sch action 6} are the ones we will couple to Schr\"odinger gravity in the main text.

Up to the order we are working ($n_t\leq 2$ and $n_s\leq 4$), it is not possible to form any additional SFT that is linearly independent from the ones we obtained above in the set of SFT$_{1-10}$. Finally, we come back to the invariant SFT$_{6'}$ that we found in eq. \eqref{eq: flat P5} and show that it is not an independent invariant. This follows from the relation:
\begin{align}\label{eq: relation P5 to other}
S^{(6')} = -d^2\mass ^2\left(S^{(6)} - S^{(7)}\right) -2d\mass S^{(10)} + (d-1)S^{(4)} + (d+2)S^{(3)} - S^{(2)} \,.
\end{align}
This is the reason why we added a prime on its label. We do not couple the SFT$_{6'}$ in the main text where we choose to work with the set of SFT$_{1-10}$ constructed from $\Psi$ and $\varphi$.

\providecommand{\href}[2]{#2}\begingroup\raggedright\endgroup


\end{document}